# Isotropic forms of dynamics in the relativistic direct interaction theory

A.Duviryak, V.Shpytko, V.Tretyak

Institute for Condensed Matter Physics of
the National Academy of Sciences of Ukraine
1 Svientsitskii St., 290011 Lviv–11, Ukraine

The Lagrangian relativistic direct interaction theory in the various forms of dynamics is formulated and its connections with the Fokker-type action theory and with the constrained Hamiltonian mechanics are established. The motion of classical two-particle system with relativistic direct interaction is analysed within the framework of isotropic forms of dynamics in the two- and four-dimensional space-time. Some relativistic exactly solvable quantum-mechanical models are also discussed.

## 1. Introduction

The relativistic direct interaction theory arises from the expectation that the dynamics of an interacting particle system can be constructed in a consistent Poincaré-invariant way without introducing the notion of the field as an independent object with its own degrees of freedom [ 1, 2, 3, 4, 5]. At present the principal possibility of such a theory is evident in the classical and quantum domains. Its application to the description of particle systems is most effective when processes of radiation and particle creation may be neglected.

Among various more or less equivalent approaches to the construction of the relativistic direct interaction theory, the single-time Lagrangian formalism [ 1, 6, 7] proposed by Professor Gaida more than twenty years ago, seems to be the most convenient for the consideration of the general problem of relativistic dynamics, as well as for the investigation of various approximations. This formalism has been extended to an arbitrary form of relativistic dynamics [ 8] defined geometrically by means of space-like foliations of the Minkowski space [ 9, 7, 10]. The conditions of the Poincaré-invariance were reformulated in an arbitrary form of dynamics and a wide class of exact solutions to the equations expressing these conditions were established for the interactions originally described by a Fokker-type action. The transition from the classical Lagrangian to the Hamiltonian description allows one to consider the relativistic effects in the statistical and quantum mechanical properties of the particle systems.

The purpose of the present paper is to review some relatively recent generalizations and specifications of this development. The transition from a non-relativistic interacting particle system to its relativistic counterpart, which on a more formal level can be considered as the replacement of the Galilei group by the Poincaré group as a symmetry group of the system, leads to profound changes in the structure of the theory. Within the Lagrangian formalism such a change manifests itself in the necessity of using the interaction Lagrangians depending on derivatives of an infinitely high order: in the general case the exact relativistic Lagrangian must be defined on the infinite order jet space [ 6]. This fact is the Lagrangian counterpart of the famous no-interaction theorem in the Hamiltonian relativistic mechanics [ 11] and has with the latter a common cause lying in the very structure of the Poincaré group. It also reflects the time non-locality inherent to relativistic interactions. All the aforementioned exact solutions of Poincaré-invariance conditions corresponding to time-symmetric Fokker-type actions have such kind of non-locality in any form of relativistic mechanics [ 12, 13, 10]. Although there are elaborated several methods of dealing with such systems (expansions in various parameters [ 1, 14, 15], transition to the center-of-mass variables [ 16]), it is evident that such a drastic change in the structure of mechanical description leads to serious difficulties in the physical interpretation of the formalism, as well as in proving its mathematical consistency.

But there are important exceptions from the general rule. If the form of dynamics defines a simultaneity relation in the Poincaré-invariant way (i.e. the Poincaré group transforms simultaneous events into simultaneous ones), then the corresponding invariance conditions of the Lagrangian description allow a large class of exact solutions containing derivatives of any finite order (not less than unity). Particularly, in such forms of dynamics we can construct in the closed form a variety of nontrivial interaction Lagrangians depending on the first order derivatives.

This fact was first established for an $N$-particle system in the two-dimensional space-time $\mathbb{M}_2$ within the framework of the front form of dynamics [ 17]. Then it was extended to the case of a two-particle system in the four-dimensional Minkowski space $\mathbb{M}_4$ by means of isotropic forms of dynamics with simultaneity between the events of particle world lines defined by a light cone [ 18]. The existence of such "standard" relativistic Lagrangians brings the problem of describing such kind of systems within the scope of the usual analytical (and, probably, quantum) mechanics. It allows the formulation of various exact models of relativistic direct interactions which admit more or less explicit investigations. Such models are the main subject of this paper.

It is organized as follows. In section 1 we begin with introducing the notion of the form of relativistic dynamics within the framework of the Lagrangian formalism. The general features of the relativistic Lagrangian description in a two-dimensional version of the front form of dynamics and in isotropic forms of dynamics are discussed in sections 3 and 4, respectively. The Fokker-type action integrals which correspond to time-asymmetric interactions are considered in section 5. Section 6 is devoted to the construction of the Hamiltonian formalism with constraints in the isotropic form of dynamics. On this basis in section 7 we investigate in the most explicit form the motions of two particles under the influence of a time-asymmetric scalar, vector, and other interactions of physical interest. The limiting case of straight line motions of such systems is considered in section 8 within the framework of the front form of dynamics. Finally, in section 9 we present certain exactly solvable relativistic quantum models of interacting particle systems in the two-dimensional space-time.

## 2. Geometrical concept of the forms of dynamics

Let us consider a dynamical system consisting of $N$ interacting point particles. It is convenient to describe the evolution of this system in the 4-dimensional Minkowski space $\mathbb{M}_4$ with coordinates $x^\mu$, $\mu = 0, 1, 2, 3$. We use the metric $\|\eta_{\mu\nu}\| = \mathrm{diag}(1, -1, -1, -1)$. The motion of the particles is described by the world lines $\gamma_a : \mathbb{R} \to \mathbb{M}_4$, $a = 1, ..., N$, which can be parametrized by arbitrary parameters $\tau_a$. In the coordinates we have

$$\gamma_a : \tau_a \mapsto x_a^\mu(\tau_a). \tag{2.1}$$

The velocity of light is taken to be unity.

Since in the Poincaré-invariant theory no particle can move with the velocity greater than the velocity of light, the world lines $\gamma_a$ must be time-like lines, and the tangent vectors

$$u_a^\mu = \frac{\mathrm{d}x_a^\mu}{\mathrm{d}\tau_a} \tag{2.2}$$

obey the inequality

$$u_a^2 \equiv \eta_{\mu\nu} u_a^\mu u_a^\nu \equiv u_a \cdot u_a > 0. \tag{2.3}$$

It is well known that the whole physical information about the motion of the system is contained in the world lines $\gamma_a$ considered as unparametrized paths in the Minkowski space. Therefore, freedom in the choice of parameters $\tau_a$ may be used for the simplification of the description. Particularly, we can choose common parameter $t$ for all the world lines of the $N$-particle system. This parameter is defined by a set of $N$ relations of the following general form:

$$\Phi_a(x_1(t), \ldots, x_N(t), u_1(t), \ldots, u_N(t), t) = 0. \tag{2.4}$$

The geometrical concept of the forms of relativistic dynamics originated by Dirac [ 8, 19] can be introduced within the framework of the single-time Lagrangian or Hamiltonian descriptions in the following way [ 9, 7, 10]. Let us consider the foliation $\Sigma = \{\Sigma_t | t \in \mathbb{R}\}$ of the Minkowski space $\mathbb{M}_4$ by the hypersurfaces $\Sigma_t$ with the equation

$$t = \sigma(x), \qquad t \in \mathbb{R}. \tag{2.5}$$

We require that every hypersurface $\Sigma_t = \{x \in \mathbb{M}_{n+1} | \sigma(x) = t\}$ must intersect the world lines $\gamma_a$ of all the particles at one and only one point

$$x_a(t) = \gamma_a \bigcap \Sigma_t. \tag{2.6}$$

This allows us to consider $t$ as an evolution parameter of the system [ 19, 20]. In the Poincaré-invariant theory, when we consider only time-like world lines, the hypersurfaces (2.5) must be space-like or isotropic,

$$\eta_{\mu\nu}(\partial^\mu \sigma)(\partial^\nu \sigma) \geq 0, \tag{2.7}$$

where $\partial^\mu = \partial/\partial x_\mu$. Then we have $\partial_0 \sigma > 0$, and the hypersurface equation (2.5) has the unique solution $x^0 = \varphi(t, \mathbf{x})$, where $\mathbf{x} = (x^i)$, $i = 1, 2, 3$. Therefore, the constraint $x_a(t) \in \Sigma_t$ enables us to determine the zeroth component of $x_a(t)$ in terms of $t$ and $x_a^i(t)$. The parametric equations (2.1) of the world lines of the particles in the given form of dynamics are as follows:

$$x^0 = \varphi(t, \mathbf{x}_a(t)) \equiv \varphi_a, \qquad x^i = x_a^i(t). \tag{2.8}$$

The evolution of the system is determined by $3N$ functions $t \mapsto x_a^i(t)$. They may be considered as representatives (in some local chart) for the sections $s : \mathbb{R} \to \mathbb{F}, t \mapsto (t, x_a^i(t))$ of the trivial fibre bundle $\pi : \mathbb{F} \to \mathbb{R}$ with $3N$-dimensional fibre space $\mathcal{M} = \mathbb{R}^{3N}$ [ 21]. The latter constitutes the configuration space of our system.

Three Dirac forms of relativistic dynamics correspond to the following hypersurfaces (2.5): $x^0 = t$ (instant form), $x^0 - x^3 = t$ or $x^0 + x^3 = t$ (front form), and $\eta_{\mu\nu} x^\mu x^\nu = t^2$ (point form). Other examples may be found in [ 9].

Now we assume that the evolution of the system under consideration is completely determined by specifying the action functional

$$S = \int \mathrm{d}t L. \tag{2.9}$$

The Lagrangian function $L : J^\infty \pi \to \mathbb{R}$ is defined on the infinite order jet space of the fibre bundle $\pi : \mathbb{F} \to \mathbb{R}$ with the standard coordinates $x_a^{i(s)}$ [ 22, 23]. The values of these coordinates for the section $s : t \mapsto (t, x_a^i(t))$ belonging to the corresponding equivalence class from $J^\infty \pi$ are $x_a^{i(s)}(t) = \mathrm{d}^s x_a^i(t)/\mathrm{d}t^s$, $s = 0, 1, 2, \ldots$. The variational principle $\delta S = 0$ with the action (2.9) gives Euler-Lagrange equations of motion

$$\sum_{s=0}^{\infty} (-D)^s \frac{\partial L}{\partial x_a^{i(s)}} = 0, \tag{2.10}$$

where $D$ is an operator of the total time derivative

$$D = \sum_a \sum_{s=0}^{\infty} x_a^{i(s+1)} \frac{\partial}{\partial x_a^{i(s)}} + \frac{\partial}{\partial t}. \tag{2.11}$$

Let us consider an arbitrary $r$-parametric Lie group $\mathcal{G}$ acting on $\mathbb{M}_4$ by the point transformations $g : \mathbb{M}_4 \to \mathbb{M}_4$:

$$x^\mu \mapsto (gx)^\mu = x^\mu + \lambda^\alpha \zeta_\alpha^\mu(x) + o(\lambda), \tag{2.12}$$

where $\lambda^\alpha$, $\alpha = 1, \ldots, r$ are the parameters of the group. The vector fields

$$\mathcal{X}_\alpha = \zeta_\alpha^\mu \partial_\mu \tag{2.13}$$

satisfy the commutation relations of the Lie algebra of group $\mathcal{G}$,

$$[\mathcal{X}_\alpha, \mathcal{X}_\beta] = c_{\alpha\beta}^\gamma \mathcal{X}_\gamma, \qquad \alpha, \beta, \gamma = 1, \ldots, r, \tag{2.14}$$

with the structure constants $c_{\alpha\beta}^\gamma$.

The action (2.12) of group $\mathcal{G}$ on $\mathbb{M}_4$ can be easily extended on the world lines $\gamma_a$ by the rule:

$$\gamma_a \mapsto g\gamma_a = \{gx | x \in \operatorname{Im}\gamma_a\}. \tag{2.15}$$

But in the given form of dynamics the world lines $\gamma_a$ are determined by the functions $t \mapsto x_a^i(t)$ or, in other words, by sections $s$ of the bundle $\pi$. Therefore, (2.15) induces an action of group $G$ on $J^\infty\pi$ by the Lie-Bäcklund transformations [ 22, 24, 23]. As it was shown in [ 9], the generators of such transformations have the form:

$$X_\alpha = \sum_a \sum_{s=0}^\infty (D^s \xi_{a\alpha}^i) \frac{\partial}{\partial x_a^{i(s)}}, \tag{2.16}$$

where

$$\xi_{a\alpha}^i = \zeta_{a\alpha}^i - v_a^i \eta_{a\alpha}, \tag{2.17}$$

and

$$\zeta_{a\alpha}^i = \zeta_\alpha^i(t, \mathbf{x_a}), \qquad \eta_{a\alpha} = (X_\alpha \sigma)(t, \mathbf{x_a}), \qquad v_a^i = x_a^{i(1)}. \tag{2.18}$$

The Lie-Bäcklund vector fields (2.16) obey the same commutation relations as (2.14),

$$[X_\alpha, X_\beta] = c_{\alpha\beta}^\gamma X_\gamma, \tag{2.19}$$

and commute with the total time derivative (2.11)

$$[X_\alpha, D] = 0. \tag{2.20}$$

For the Poincaré group we have the following ten vector fields corresponding to the natural action of $\mathcal{P}(1, 3)$ on $\mathbb{M}_4$:

$$\mathcal{X}_\mu^T = \partial_\mu, \tag{2.21}$$

$$\mathcal{X}_{\mu\nu}^L = x_\mu \partial_\nu - x_\nu \partial_\mu, \tag{2.22}$$

with the commutation relations

$$[\mathcal{X}_\mu^T, \mathcal{X}_\mu^T] = 0, \tag{2.23}$$

$$[\mathcal{X}_\mu^T, \mathcal{X}_{\rho\sigma}^L] = \eta_{\mu\rho}\mathcal{X}_\sigma^T - \eta_{\mu\sigma}\mathcal{X}_\rho^T, \tag{2.24}$$

$$[\mathcal{X}_{\mu\nu}^L, \mathcal{X}_{\rho\sigma}^L] = \eta_{\nu\rho}\mathcal{X}_{\mu\sigma}^L + \eta_{\mu\sigma}\mathcal{X}_{\nu\rho}^L - \eta_{\mu\rho}\mathcal{X}_{\nu\sigma}^L - \eta_{\nu\sigma}\mathcal{X}_{\mu\rho}^L. \tag{2.25}$$

Thus, we obtain the following realization of the Poincaré algebra in terms of Lie-Bäcklund vector fields (2.16):

$$X_\mu^T = \sum_a \sum_{s=0}^\infty D^s[\delta_\mu^i - v_a^i\sigma_{a\mu}]\frac{\partial}{\partial x_a^{i(s)}}, \tag{2.26}$$

$$X_{\mu\nu}^L = \sum_a \sum_{s=0}^\infty D^s[x_{a\mu}\delta_\nu^i - x_{a\nu}\delta_\mu^i - v_a^i(x_{a\mu}\sigma_{a\nu} - x_{a\nu}\sigma_{a\mu})]\frac{\partial}{\partial x_a^{i(s)}}, \tag{2.27}$$

where we must use (2.8) for the elimination of $x_a^0$, and we denote

$$\sigma_{a\mu} \equiv (\partial_\mu\sigma)(t, \mathbf{x}_a). \tag{2.28}$$

Making use of the hypersurface equation (2.5) we find:

$$\sigma_{a0} = (\partial\varphi_a/\partial t)^{-1} \equiv \varphi_{at}^{-1}, \tag{2.29}$$

$$\sigma_{ai} = -\varphi_{at}^{-1}(\partial\varphi_a/\partial x_{ai}) \equiv -\varphi_{at}^{-1}\varphi_{ai}. \tag{2.30}$$

It is convenient to introduce the vector fields

$$\mathcal{H} = -X_0^T, \quad \mathcal{P}_i = X_i^T, \quad \mathcal{J}_i = -\frac{1}{2}\varepsilon_{ijk}X_{jk}^L, \quad \mathcal{K} = X_{i0}^L, \tag{2.31}$$

obeying the following commutation relations:

$$[\mathcal{H}, \mathcal{P}_i] = 0, \qquad [\mathcal{P}_i, \mathcal{P}_j] = 0, \qquad [\mathcal{H}, \mathcal{J}_i] = 0, \qquad [\mathcal{P}_i, \mathcal{J}_k] = -\varepsilon_{ikl}\mathcal{P}_l, \tag{2.32}$$

$$[\mathcal{J}_i, \mathcal{J}_k] = -\varepsilon_{ikl}\mathcal{J}_l, \qquad [\mathcal{K}_i, \mathcal{J}_k] = -\varepsilon_{ikl}\mathcal{K}_l, \qquad [\mathcal{K}_i, \mathcal{K}_j] = \varepsilon_{ijk}\mathcal{J}_k, \tag{2.33}$$

$$[\mathcal{H}, \mathcal{K}_i] = \mathcal{P}_i, \qquad [\mathcal{P}_i, \mathcal{K}_j] = \delta_{ij}\mathcal{H}. \tag{2.34}$$

Inserting (2.29), (2.30) into (2.26), (2.27), we obtain the realization of the Poincaré algebra which is convenient for the consideration of the symmetries of a single-time three-dimensional Lagrangian description [ 9]:

$$\mathcal{H} = \sum_a \sum_{s=0}^\infty D^s[v_a^i\varphi_{at}^{-1}]\frac{\partial}{\partial x_a^{i(s)}}, \tag{2.35}$$

$$\mathcal{P}_i = \sum_a \sum_{s=0}^\infty D^s[\delta_i^j + v_a^j\varphi_{ai}\varphi_{at}^{-1}]\frac{\partial}{\partial x_a^{j(s)}}, \tag{2.36}$$

$$\mathcal{J}_i = \varepsilon_{ikl} \sum_a \sum_{s=0}^\infty D^s[x_a^k(\delta_l^j + v_a^j\varphi_{al}\varphi_{at}^{-1})]\frac{\partial}{\partial x_a^{j(s)}}, \tag{2.37}$$

$$\mathcal{K}_i = \sum_a \sum_{s=0}^\infty D^s[-\varphi_a\delta_i^j + v_a^j(x_{ai} - \varphi_a\varphi_{ai})\varphi_{at}^{-1}]\frac{\partial}{\partial x_a^{j(s)}}. \tag{2.38}$$

The symmetry of the Lagrangian description of an interacting particle system under group $\mathcal{G}$ means the invariance of the Euler-Lagrange equation (2.10) under corresponding Lie-Bäcklund transformations generated by the vector fields (2.16). The sufficient conditions for the symmetry under the Poincaré group have the form [6, 7]:

$$X_\alpha L = D\Omega_\alpha, \qquad \alpha = 1, \ldots, 10, \tag{2.39}$$

with auxiliary functions $\Omega_\alpha$, satisfying the consistency relations

$$X_\alpha \Omega_\beta - X_\beta \Omega_\alpha = c^\gamma_{\alpha\beta} \Omega_\gamma. \tag{2.40}$$

An important corollary of symmetry conditions (2.39), (2.40) for an arbitrary $r$-parametric Lie group is the existence of $r$ conservation laws

$$DG_\alpha = 0, \qquad \alpha = 1, \ldots, r, \tag{2.41}$$

for quantities $G_\alpha$ which can be explicitly determined in terms of the Lagrangian function $L$ and auxiliary functions $\Omega_\alpha$. This statement, which is well known as the Nöther theorem, follows immediately from the identity [22, 24]

$$X_\alpha L = \sum_a \xi^i_{a\alpha} \mathcal{E}_{ai} L + D \sum_a \sum_{s=o}^\infty \pi_{ai,s} D^s \xi^i_{a\alpha}, \tag{2.42}$$

which holds for an arbitrary Lie-Bäcklund vector field (2.16). Here,

$$\pi_{ai,s} = \sum_{n=s}^\infty (-D)^{n-s} \frac{\partial L}{\partial x_a^{i(n+1)}} \tag{2.43}$$

are the Ostrogradskyj momenta. Making use of the identity (2.42) in symmetry conditions (2.39), one readily checks that for the solutions of Euler-Lagrange equation (2.10) the conservation laws (2.41) hold with

$$G_\alpha = \sum_a \sum_{s=o}^\infty \pi_{ai,s} D^s \xi^i_{a\alpha} - \Omega_\alpha. \tag{2.44}$$

In the general case the Poincaré-invariance conditions forbid the existence of interaction Lagrangians which are defined on the jet-space $J^r\pi$ with some finite $r$ (for example, with $r = 1$). This leads to serious difficulties in the physical interpretation of the formalism, and, in fact, makes it impossible to obtain a closed form of the corresponding Hamiltonian functions.

In the following we shall consider some exceptions from this rule. The first is offered by the front form of dynamics in the two-dimensional Minkowski space. In this case there exists a wide class of interaction Lagrangians for an $N$-particle system, which are defined on the first-order jet-space $J^1\pi$ [17]. The second consists in the consideration of a more general definition of the form of dynamics, than (2.5).

## 3. Front form of dynamics in $\mathbb{M}_2$

In the two-dimensional space-time $\mathbb{M}_2$ the front form of dynamics corresponds to the foliation of $\mathbb{M}_2$ by isotropic hyperplanes (i.e., lines):

$$x^0 + x = t. \tag{3.1}$$

In this form of dynamics for an $N$-particle system Poincaré-invariance conditions allow the existence of interaction Lagrangians which do not contain derivatives higher than the first order. The general form of such a Lagrangian function including only pairwise interactions is given by [ 17]:

$$L = -\sum_a m_a k_a + \sum_a \sum_{a \, < \, b} r_{ab} V_{ab}(r_{ab} k_a^{-1}, r_{ab} k_b^{-1}), \tag{3.2}$$

where $k_a = \sqrt{1 - 2v_a}$, $r_{ab} \equiv x_a - x_b$, $a, b = \overline{1, N}$, and $V_{ab}$ are arbitrary functions of the indicated arguments. As a result of the Poincaré invariance of the Lagrangian function (3.2), there exist three conserved quantities: energy $E$, total momentum $P$, and the center-of-inertia integral of motion $K$. They have the form [ 17]:

$$
\begin{aligned}
E &= \sum_{a=1}^{N} v_a \frac{\partial L}{\partial v_a} - L, \quad P = \sum_{a=1}^{N} \frac{\partial L}{\partial v_a} - E, \\
K &= -tP + \sum_{a=1}^{N} x_a \frac{\partial L}{\partial v_a}.
\end{aligned} \tag{3.3}
$$

The existence of the interaction Lagrangians (3.2) permits one to trace quite easily the relations between various formalisms of relativistic dynamics and to find out special features of relativistic particle systems. In spite of the fact that the Lagrangian function (3.2) does not contain higher derivatives and the transition to the Hamiltonian description is a usual Legendre transformation, the investigation of exactly solvable models shows some new features which do not occur in the non-relativistic mechanics.

In the classical mechanics, the Lagrangian function is determined on the tangent bundle $T\mathcal{M}$, $L : T\mathcal{M} \to \mathbb{R}$ [ 21]. If the configuration space $\mathcal{M}$ is diffeomorphic to $\mathbb{R}^N$, then the tangent bundle is a trivial one: $T\mathcal{M} = \mathbb{R}^N \times \mathbb{R}^N$. This means that a single chart with coordinates $(x_1, ..., x_N, v_1, ..., v_N)$ covers the whole $T\mathcal{M}$.

For the Lagrangian (3.2) the configuration space $\mathcal{M}$ coincides with the whole $\mathbb{R}^N$ or at least with the disconnected union of open sets in $\mathbb{R}^N$. Hence, one can expect that there should not be any complications connected with the global structure. But the Lagrangian function (3.2) is determined on submanifold $\mathcal{Q}_f$ of $T\mathcal{M}$. This submanifold is defined by the inequalities

$$v_a < 1/2, \tag{3.4}$$

which reflect the time-like character of particle world lines in $\mathbb{M}_2$. Submanifold $\mathcal{Q}_f$ does not have the structure of a tangent bundle.

Moreover, we do restrict the Lagrangian description to the smaller region than $T\mathcal{M}$ for another reason. The Hamilton principle $\delta S = 0$ leads to Euler-Lagrange equations if the Hessian is positively defined:

$$\mathsf{h} = \det||\partial^2 L/\partial v_a \partial v_b|| > 0. \tag{3.5}$$

For the Lagrangian function (3.2) the Hessian is, in general, a complicated function on coordinate differences and velocities: $\mathsf{h} = \mathsf{h}(r_{ab}, k_c)$. Therefore, inequality (3.5) defines an open region $\mathcal{Q} \subset TM \approx \mathbb{R}^{2N}$. This region also does not have the structure of a tangent bundle and for a free-particle system coincides with $\mathcal{Q}_f$.

It could be unimportant if the system moves inside the region (3.5) and does not reach the boundary

$$\partial\mathcal{Q} = \{(x_a, v_a) \in T\mathcal{M}|\mathsf{h} = 0, \mathsf{h}^{-1} = 0\}. \tag{3.6}$$

In contrast, the difficulty arises when the system reaches the points of the boundary region (singular points) at a finite value of the evolution parameter $t$ [25]. The theorem of existence and uniqueness for Euler-Lagrange differential equations breaks at singular points and the Lagrangian system is not defined. Therefore, we cannot prolong the evolution of the system beyond the critical points within the framework of the basic Lagrangian description.

The way of overcoming this difficulty is offered by the Hamiltonian description. It is well known that the Legendre transformation is a differentiable mapping $\mathcal{L} : T\mathcal{M} \to T^*\mathcal{M}$. The transition from the Lagrangian (3.2) to the Hamiltonian formalism may be performed by the usual Legendre transformation. But this transformation is a diffeomorphism only in the region $\mathcal{Q}$. It maps the open region $\mathcal{Q} \subset \mathbb{R}^{2N}$ to the open one $\mathcal{L}\mathcal{Q} \subset T^*\mathcal{M} \approx \mathbb{R}^4$. The Hamiltonian description is equivalent to the Lagrangian one only in the region $\mathcal{L}\mathcal{Q}$ [21]. In a strict sense the motion in the Hamiltonian case is well defined on $\mathcal{L}\mathcal{Q}$ only. In other words, we should consider $\mathcal{L}\mathcal{Q}$ as a whole phase space of the system.

After the Legendre transformation is performed, the conserved quantities (3.3) become canonical generators of the Poincaré group $\mathcal{P}(1, 1)$:

$$P_+ = \sum_{a=1}^{N} p_a, \qquad K = \sum_{a=1}^{N} x_a p_a, \tag{3.7}$$

$$P_- = \sum_{a=1}^{N} \frac{m_a^2}{p_a} + \frac{1}{P_+} V(r p_b, r_{1c}/r) . \tag{3.8}$$

They satisfy the following Poisson bracket relations:

$$\{P_+, P_-\} = 0, \qquad \{K, P_\pm\} = \pm P_\pm. \tag{3.9}$$

Here we have introduced more convenient in the front form quantities $P_\pm = E \pm P$. The classical total mass squared function $M^2 = P_+ P_-$ has vanishing Poisson brackets with all the generators (3.7), (3.8).

If we deal with the Lagrangian region $\mathcal{LQ}$ within the Hamiltonian description, we shall obtain the same results as in the Lagrangian case. For systems which reach the points of $\partial \mathcal{Q}$, the Lagrangian description leads to disconnected segments of world lines [ 26]. To obtain the whole evolution of such systems we have to determine the motion of the system beyond the Lagrangian region. In the following we shall demonstrate for certain relativistic models that the Hamiltonian formalism permits one to prolong the evolution of the system beyond singular points and obtain smooth world lines in $\mathbb{M}_2$, as well as in the four-dimensional space-time $\mathbb{M}_4$ (see sections 8 and 7.1, respectively).

## 4. Isotropic forms of dynamics

For a two-particle system in $\mathbb{M}_4$ the class of isotropic forms of dynamics corresponds to the following definition of simultaneity between the events of particle world lines [ 18]:

$$[x_1(t) - x_2(t)]^2 = 0 \tag{4.1}$$

with the supplementary condition

$$\mathrm{sgn}[x_1^0(t) - x_2^0(t)] = \epsilon, \tag{4.2}$$

where $\epsilon = \pm 1$. Such a description has been developed within the framework of the predictive relativistic mechanics in a series of papers by Künzle [ 27, 28, 29]. The idea of this definition of simultaneity was formulated in the classic Van Dam-Wigner's work [ 30]. In the contents of relativistic Lagrangian and Hamiltonian mechanics the descriptions based on equation (4.1) were elaborated in [ 18, 31, 32].

equations (4.1), (4.2) determine the difference of the zeroth components:

$$x_1^0(t) - x_2^0(t) = \epsilon |\mathbf{x}_1(t) - \mathbf{x}_2(t)|. \tag{4.3}$$

For the definition of the value of the common evolution parameter $t$ we choose the relation

$$\sigma\left(\frac{x_1(t) + x_2(t)}{2}\right) = t, \tag{4.4}$$

where $\sigma(x)$ is the same function as in the definition of the geometrical forms of dynamics (2.5). Therefore, we have

$$\frac{x_1^0(t) + x_2^0(t)}{2} = \varphi\left(t, \frac{\mathbf{x}_1(t) + \mathbf{x}_2(t)}{2}\right), \tag{4.5}$$

and

$$x_1^0 = \varphi(t, \mathbf{y}) + \frac{1}{2}\epsilon|\mathbf{r}|, \quad x_2^0 = \varphi(t, \mathbf{y}) - \frac{1}{2}\epsilon|\mathbf{r}|. \tag{4.6}$$

Here and henceforth the variables $y^\mu \equiv (x_1^\mu + x_2^\mu)/2$ and $r^\mu \equiv x_1^\mu - x_2^\mu$ are used.

If we put $\varphi(t, \mathbf{y}) = t$ as in the instant form of dynamics, we obtain

$$x_a^0 = t + \frac{1}{2}(-1)^{\bar{a}}\epsilon|\mathbf{r}|, \qquad a = 1, 2; \quad \bar{a} \equiv 3 - a. \tag{4.7}$$

These relations have been used in [ 27, 28]. When we choose $\sigma(x)$ as in the front form $[\varphi(t, \mathbf{y}) = t - y^3]$, we obtain

$$x_a^0 = t + y^3 + \frac{1}{2}(-1)^{\bar{a}}\epsilon|\mathbf{r}|. \tag{4.8}$$

In the two-dimensional space-time (4.8) reduces to the geometrical definition of the front form provided $\epsilon = \text{sgn}(x_2 - x_1)$.

The general structure of the Lagrange function is again determined by the Poincaré-invariance conditions. Their formulation requires the realization of algebra $\mathfrak{p}(1, 3)$ by the Lie-Bäcklund vector fields (2.16). In paper [ 18] it was shown, that the components of the corresponding fields have the form (2.17), where

$$\zeta_{a\alpha}^i = \zeta_\alpha^i[x_a(t)] \tag{4.9}$$

and

$$\begin{aligned}
\eta_{a\alpha} &= \frac{1}{2}[\zeta_\alpha^\nu(x_1) + \zeta_\alpha^\nu(x_2)]\partial_\nu\sigma\left(\frac{x_1 + x_2}{2}\right) \\
&= (\zeta_\alpha^\nu\partial_\nu\sigma)\left(\frac{x_1 + x_2}{2}\right) = \eta_\alpha(t, \mathbf{y}).
\end{aligned} \tag{4.10}$$

All the zeroth components here must be excluded by means of relations (4.6). Let us note the independence of $\eta_\alpha$ on the particle labels.

It is a matter of simple calculation to verify that such vector fields satisfy the commutation relations (2.19).

The Poincaré-invariance conditions have the form (2.39) where we can put

$$\Omega_\alpha = -\eta_\alpha L. \tag{4.11}$$

Such a choice of auxiliary functions $\Omega_\alpha$ enables (4.11) to be expressed in the form:

$$\hat{X}_\alpha L + LD\eta_\alpha = 0, \tag{4.12}$$

where the vector fields

$$\hat{X}_\alpha = X_\alpha + \eta_\alpha D. \tag{4.13}$$

generate the point transformation of the extended configuration space $\mathbb{F} = \mathbb{R} \times \mathcal{M}$.

As in the case of the front form of dynamics in $\mathbb{M}_2$, equations (4.12) allow a large class of exact solutions depending on the derivatives of any finite order. If we suppose that the Lagrangian contains only the first derivatives, i.e. it is defined on the space $J^1\pi$, we obtain

$$\eta_\alpha\frac{\partial L}{\partial t} + \sum_{a=1}^{2}\left(\zeta_{a\alpha}^i\frac{\partial L}{\partial x_a^i} + (D\zeta_{a\alpha}^i - v_a^iD\eta_\alpha)\frac{\partial L}{\partial v_a^i}\right) + LD\eta_\alpha = 0, \tag{4.14}$$

where $\zeta_{a\alpha}^i \equiv \zeta_\alpha^i[x_a(t)]$.

The general solution to these equations can be presented in the form [ 18]:

$$L = \vartheta F(\sigma_1, \sigma_2, \omega), \tag{4.15}$$

where

$$\vartheta = (x_1^\mu - x_2^\mu)u_{1\mu} = (x_1^\mu - x_2^\mu)u_{2\mu} = \epsilon|\mathbf{r}|D\varphi(t, \mathbf{y}) - \mathbf{r} \cdot \dot{\mathbf{y}};$$

$$\Gamma_a^{-2} = u_a^\mu u_{a\mu} = \left(D\varphi(t, \mathbf{y}) - \frac{1}{2}(-1)^a \epsilon \mathbf{n} \cdot \mathbf{v}\right)^2 - v_a^2;$$

$$\mathbf{n} \equiv \mathbf{r}/r, \qquad r \equiv |\mathbf{r}|, \qquad \mathbf{v} \equiv \mathbf{v}_1 - \mathbf{v}_2;$$

$$\sigma_a = \Gamma_a \vartheta = r^\nu \hat{u}_{a\nu}, \qquad \hat{u}_{a\nu} \equiv u_{a\nu}/\sqrt{u_a^2};$$

$$\omega = \Gamma_1 \Gamma_2 \left[(D\varphi(t, \mathbf{y}))^2 - \mathbf{v}_1 \cdot \mathbf{v}_2 - \frac{1}{4}(\mathbf{n} \cdot \mathbf{v})^2\right] = \hat{u}_{1\nu}\hat{u}_2^\nu,$$

and $F$ being an arbitrary (smooth) function on three variables.

In the front form of dynamics in $\mathbb{M}_2$ we have $\vartheta = r$, $\Gamma_a = (1 - 2v_a)^{-1/2} = k_a^{-1}$, and $\omega$ is a function on the invariants $\sigma_1$, $\sigma_2$:

$$\omega = \frac{1}{2}\left(\frac{\Gamma_1}{\Gamma_2} + \frac{\Gamma_2}{\Gamma_1}\right) = \frac{1}{2}\left(\frac{\sigma_1}{\sigma_2} + \frac{\sigma_2}{\sigma_1}\right). \tag{4.16}$$

Invariance conditions (4.12) lead to the conservation laws for the quantities (2.44). In our case they have the form:

$$G_\alpha = \sum_{a=1}^{2}(\zeta_{a\alpha}^i - v_a^i \eta_\alpha)\frac{\partial L}{\partial v_a^i} - \Omega_\alpha. \tag{4.17}$$

Taking into account (4.11) they can be expressed as

$$G_\alpha = \sum_{a=1}^{2}\zeta_{a\alpha}^i \frac{\partial L}{\partial v_a^i} - \eta_\alpha H, \tag{4.18}$$

where

$$H = \sum_{a=1}^{2} v_a^i \frac{\partial L}{\partial v_a^i} - L. \tag{4.19}$$

Let us introduce the Poincaré-invariant functions:

$$A_a = \sigma_a^2 \frac{\partial F}{\partial \sigma_a} + (\omega\sigma_a - \sigma_{\bar{a}})\frac{\partial F}{\partial \omega}, \tag{4.20}$$

$$B_a = \sigma_a^2 \frac{\partial F}{\partial \sigma_a} + (\omega\sigma_a + \sigma_{\bar{a}})\frac{\partial F}{\partial \omega}. \tag{4.21}$$

They are not independent,

$$\sigma_1(A_1 - B_1) = \sigma_2(A_2 - B_2). \tag{4.22}$$

In terms of these functions we have:

$$\frac{\partial L}{\partial v_a^i} = \frac{1}{2}(r\varphi_i(t,\mathbf{y}) - \epsilon r_i)\tilde{F} + v_{ai}\Gamma_a\sigma_a\left(\sigma_a\frac{\partial F}{\partial \sigma_a} + \omega\frac{\partial F}{\partial \omega}\right) - v_{bi}\Gamma_{\bar{a}}\sigma_a\frac{\partial F}{\partial \omega}$$

$$- \frac{1}{2}\varphi_i(t,\mathbf{y})(\Gamma_1 u_1^0 A_1 + \Gamma_2 u_2^0 A_2) + \frac{1}{2}(-1)^a \epsilon n_i(\Gamma_1 u_1^0 B_1 - \Gamma_2 u_2^0 B_2). \tag{4.23}$$

where

$$\tilde{F} = F + \sum_{a=1}^{2}\sigma_a\frac{\partial F}{\partial \sigma_a}. \tag{4.24}$$

The function (4.19) is easily found to be

$$H = \varphi_t(t,\mathbf{y})(-r\tilde{F} + \Gamma_1 u_1^0 A_1 + \Gamma_2 u_2^0 A_2). \tag{4.25}$$

Explicitly, the integrals of motion (4.18) are given by

$$G_\alpha = -\zeta_\alpha^0(t,\mathbf{y})\varphi_t(t,\mathbf{y})^{-1}H + \zeta_\alpha^i(t,\mathbf{y})(-\epsilon r_i\tilde{F} + \Gamma_1 v_{1i}A_1 + \Gamma_2 v_{2i}A_2)+$$

$$+ \frac{1}{2}(\zeta_{1\alpha}^i - \zeta_{2\alpha}^i)[\Gamma_2(\epsilon n_i u_2^0 - v_{2i})B_2 - \Gamma_1(\epsilon n_i u_1^0 - v_{1i})B_1]. \tag{4.26}$$

Inserting the expressions for functions $\zeta_\alpha^\nu$ which correspond to the generators (2.21), (2.22) of the Poincaré group, we obtain the following formulae for conserved energy $E$, momentum $\mathbf{P}$, angular momentum $\mathbf{J}$ and the center-of-inertia integral of motion $\mathbf{K}$:

$$E = \varphi_t^{-1}H = -r\tilde{F} + \Gamma_1 u_1^0 A_1 + \Gamma_2 u_2^0 A_2, \tag{4.27}$$

$$\mathbf{P} = -\epsilon\mathbf{r}\tilde{F} + \Gamma_1\mathbf{v}_1 A_1 + \Gamma_2\mathbf{v}_2 A_2, \tag{4.28}$$

$$\mathbf{J} = \mathbf{y}\times\mathbf{P} + \frac{1}{2}\mathbf{r}\times(\Gamma_1\mathbf{v}_1 B_1 - \Gamma_2\mathbf{v}_2 B_2), \tag{4.29}$$

$$\mathbf{K} = \mathbf{y}E - \varphi(t,\mathbf{y})\mathbf{P} - \frac{1}{2}[\Gamma_2(\mathbf{r}u_2^0 - \epsilon r\mathbf{v}_2)B_2 - \Gamma_1(\mathbf{r}u_1^0 - \epsilon r\mathbf{v}_1 B_1)]. \tag{4.30}$$

We note that the expressions (4.27), (4.28) can be united into a 4-vector of momentum $P_\mu$, as well as equations (4.29), (4.30) represent a 4-tensor of angular momentum $J_{\mu\nu}$

$$P_\mu = \epsilon r_\mu\tilde{F} - \hat{u}_1\mu A_1 - \hat{u}_{2\mu}A_2, \tag{4.31}$$

$$J_{\mu\nu} = \frac{1}{2}\left(y_\nu P_\mu - y_\mu P_\nu - r_\nu(\hat{u}_{1\mu}B_1 - \hat{u}_{2\mu}B_2) - r_\mu(\hat{u}_{1\nu}B_1 - \hat{u}_{2\nu}B_2)\right). \tag{4.32}$$

Here

$$E = -P_0, \quad J_i = \epsilon_{ilk}J^{lk}, \quad K_i = J_{0i}, \tag{4.33}$$

and the zeroth components of the 4-vectors $x_a$ and $\hat{u}_a$ must be excluded with the help of relations (4.6).

The structure of the motion integrals (4.31), (4.32) agrees with the results of Refs. [ 27, 28 ] which were derived within the framework of the predictive relativistic dynamics.

Ten integrals of motion can be used to reduce the integration of equations of motion to quadratures. But it is more convenient to preform such a reduction by means of the techniques of the constrained Hamiltonian mechanics.

## 5. Fokker-type action and single-time Lagrangians

One of the possible ways to specify the form of the arbitrary functions entering the general solution of the Poincaré-invariance conditions is the comparison with the Fokker-type relativistic mechanics [ 4, 5, 33], the oldest attempt to construct the relativistic direct interaction theory which has a relation to the field description. It is based on the manifestly Poincaré-invariant variational principle formulated in terms of four-dimensional coordinates and velocities of the particles. Such a variational principle was first introduced for the electromagnetic interaction by Schwarzschild, Tetrode, and Fokker at the beginning of this century and developed by various authors (see Refs. [ 1, 5, 33] and references therein). Later this description was extended to other relativistic interactions. The equations of motion following from such a variational principle explicitly satisfy the demand of relativistic invariance and can be compared with the corresponding field theory expressions. However, this approach is not free of difficulties both on physical and mathematical levels. The cost for a manifestly Poincaré-invariant four-dimensional description is the necessity to use a many-time formalism which complicates the physical interpretation of its results. Mathematically, it is hard to motivate the obtaining of the equations of motion from the action integrals which are obviously divergent because the integration is carried out on the whole length of the world lines of the particles [ 4].

Within the framework of Fokker-type mechanics the dynamics of a relativistic particle system is specified in a manifestly Poincaré- and reparametrization-invariant way on the basis of the variational principle $\delta S = 0$ with the action being given by

$$S = S_f - S_{int},    \tag{5.1}$$

where

$$S_f = -\sum_a m_a \int \mathrm{d}\tau_a \sqrt{u_a^2},    \tag{5.2}$$

corresponds to a free-particle system and

$$S_{int} = \sum_a \sum_{a<b} \int \mathrm{d}\tau_a \int \mathrm{d}\tau_b \Lambda_{ab}(x_a, x_b, u_a, u_b).    \tag{5.3}$$

determines two-particle interactions. Here $\Lambda_{ab}$ are some functions depending on the four-dimensional particle coordinates $x_a^\mu$ and on the first derivatives $u_a^\mu$. They have the form [ 33, 35]:

$$\Lambda_{ab} = \sqrt{u_a^2 u_b^2}\, U_{ab}(x_a, x_b, \hat{u}_a, \hat{u}_b),    \tag{5.4}$$

where $\hat{u}_a^\mu = u_a^\mu / \sqrt{u_a^2}$ and function $U_{ab}$ (which we shall call the *Fokker potential*) depends on the following set of the two-body Lorentz scalars:

$$\varrho_{ab} = (x_a - x_b)^2, \qquad \sigma_{ab} = (x_a - x_b) \cdot \hat{u}_a, \qquad \omega_{ab} = \hat{u}_a \cdot \hat{u}_b;    \tag{5.5}$$

that is

$$U_{ab} = U_{ab}(\varrho_{ab}, \sigma_{ab}, \sigma_{ba}, \omega_{ab}).    \tag{5.6}$$

In papers [ 12, 13] it was shown that many-time Fokker-type action integrals can be transformed into single-time actions with non-local Lagrangians depending on the three-dimensional coordinates of the particles and on all the derivatives of the coordinates with respect to parameter $t$. Such Lagrangians provide us with a useful tool for the consideration of various approximations [ 12, 13, 7], as well as for the transition to the predictive relativistic mechanics and Hamiltonian formalism [ 14, 15]. It was demonstrated [ 13] that non-local Lagrangians corresponding to the manifestly Poincaré-invariant action integrals satisfy the Poincaré-invariance conditions within the framework of the three-dimensional Lagrangian description of interacting particle systems [ 6]. The conservation laws which follow from such an invariance were investigated via the Nöther theorem. Moreover, the non-local single-time Lagrangians which are found on the basis of the Fokker-type action integrals represent a closed form for a wide class of solutions of equations (2.29) expressing the requirements of the invariance of the Lagrangian description of particle systems under the Poincaré group [ 13, 10].

If $U_{ab}$ happens to have the special form

$$U_{ab} = e_a e_b \omega_{ab} \delta(\varrho_{ab}),$$ (5.7)

then action (5.1) describes the electromagnetic interaction of charges $e_a$ within the framework of the Tetrode-Fokker-Wheeler-Feynman electrodynamics. Such an approach has been extended to the interactions which are mediated by massive scalar and vector fields [ 4, 33, 35]:

the scalar case $\qquad U_{ab} = g_a g_b G^{sym}(\varrho_{ab}),$ (5.8)

the vector case $\qquad U_{ab} = g_a g_b \omega_{ab} G^{sym}(\varrho_{ab}).$ (5.9)

In the above, $g_a$ is a coupling constant of particle $a$ and $G^{sym}(x) = G^{sym}(x^2)$ is a time-symmetric Green function of the Klein-Gordon equation

$$(\Box + \kappa^2)G^{sym}(x) = 4\pi\delta(x),$$ (5.10)

where $\Box \equiv \eta_{\mu\nu}\partial^\mu\partial^\nu$ and $\kappa$ is a mass of the field quanta. Explicitly, we have

$$G^{sym}(x) = \delta(x^2) - \Theta(x^2)\frac{\kappa}{2\sqrt{x^2}}J_1(\kappa\sqrt{x^2}),$$ (5.11)

where $\Theta(x)$ is the Heaviside step function and $J_1(x)$ is the Bessel function of order 1.

There exists a wider class of physically important Fokker-type integrals which permit a field-theoretical interpretation of interaction between particles. It corresponds to Fokker potentials of the following form:

$$U_{ab} = g_a g_b f(\omega_{ab})G(\varrho_{ab}),$$ (5.12)

where $f(\omega)$ depends upon the tensor structure of the fields mediating the interaction, and $G(x)$ is a symmetrical Green function of the relevant wave equation. In the case of massless fields $G(x) = \delta(x^2)$. Especially, for interactions mediated by the massless field with the given helicity $\lambda = \pm n$ we have [ 36]

$$f(\omega) = T_n(\omega). \tag{5.13}$$

One more example is a model of confinement interaction [ 37], for which

$$U_{ab} = g_a g_b \sigma_{ab} \sigma_{ba} \delta(\varrho_{ab}). \tag{5.14}$$

Equivalently, this model can be presented in the form (5.12) with $f(\omega) = \omega$ and the Green function $G(x)$ replaced by the "phenomenological propagator" $\Theta(x^2)$.

Generally, the Fokker-type action with a time-symmetric Green function leads to non-local in time Lagrangians and integral- or difference-differential equations of motion. It makes the analysis of particle motions a complicated task (except for the case of circular motion when the solution may be constructed explicitly [ 38, 39]). An interesting possibility to obtain ordinary differential equations of motion is to replace $G$ in the right-hand side of equation (5.12) by the retarded (advanced) Green function of d'Alambert equation [ 34]:

$$G_\epsilon(x) = 2\Theta(\epsilon x^0)\delta(x^2), \qquad \epsilon = \pm 1. \tag{5.15}$$

This choice in the case of a two-particle system corresponds to the model with the following particle interaction: the advanced field of the first particle acts on the second particle and the retarded field of the second particle acts on the first particle. Such interactions correspond to the exact solutions of the Poincaré-invariance conditions considered above in the front and isotropic forms of dynamics [ 17, 18, 32].

In such models a one-to-one correspondence of points of two particle world lines appears naturally, namely, of those points which satisfy the *light cone condition*:

$$r^2 = 0, \quad \epsilon r^0 > 0, \qquad \text{i.e.,} \qquad \epsilon r^0 = |\mathbf{r}|, \tag{5.16}$$

This correspondence allows one to reduce the Fokker-type integral to a manifestly covariant single-time action,

$$S_I = \int d\tau \, (L + \lambda r^2), \tag{5.17}$$

where the Lagrangian multiplier $\lambda$ is introduced to take into account condition (5.16) as a holonomic constraint (the boundary constraint $\epsilon r^0 > 0$ is also meant).

An action of this kind occurs when the Fokker potential has a more general structure:

$$U = \tilde{f}(\omega, \sigma_1, \sigma_2) G_\epsilon(r), \qquad \sigma_1 \equiv \sigma_{12}, \quad \sigma_2 \equiv \sigma_{21}. \tag{5.18}$$

The relevant Lagrangian function reads:

$$L = -\sum_{a=1}^{2} m_a \sqrt{\dot{x}_a^2} - \frac{\sqrt{\dot{x}_a^2}\sqrt{\dot{x}_b^2}}{|\dot{y} \cdot r|} \tilde{f}, \tag{5.19}$$

where the dot denotes a derivative on parameter $\tau$. It creates a sufficiently wide class of two-particle *time-asymmetric* models. Their study would not be successful without an appropriate Hamiltonian description.

## 6. Hamiltonian description in the isotropic form of dynamics

The Lagrangian description in the configuration space $\mathbb{M}_4^2$ allows a natural transition to the manifestly covariant Hamiltonian description with constraints [ 40, 34]. The corresponding phase space $T^*\mathbb{M}_4^2$ is a 16-dimensional one with the Poisson brackets [..., ...]. They have a standard form in terms of covariant coordinates $x_a^\mu$ and conjugated momenta defined in a usual manner:

$$p_{a\mu} = \partial L / \partial \dot{x}_a^\mu. \tag{6.1}$$

Since the Lagrangian (5.19) and the constraint (5.16) are Poincaré-invariant, there exist ten Nöther integrals of motion,

$$P_\mu = \sum_{a=1}^2 p_{a\mu}, \qquad J_{\mu\nu} = \sum_{a=1}^2 \left( x_{a\mu} p_{a\nu} - x_{a\nu} p_{a\mu} \right). \tag{6.2}$$

In the Hamiltonian description these $P_\mu$ and $J_{\mu\nu}$ are generators of the canonical realization of the Poincaré group.

By virtue of parametric invariance of the action (5.17), the Lagrangian (5.19) is singular. Hence the canonical Hamiltonian vanishes, while the dynamics of the system is determined by the *dynamical* constraint of the following general form:

$$\phi(P^2, \ p_\perp^2, \ P \cdot r, \ p_\perp \cdot r) = 0, \tag{6.3}$$

which appears together with the holonomic constraint (5.16); here $p_{\perp\mu} \equiv p_\mu - r_\mu P \cdot p / P \cdot r$; $P_\mu$ and $p_\mu = \frac{1}{2}(p_{1\mu} - p_{2\mu})$ are canonical momenta conjugated to $y^\mu$ i $r^\mu$, respectively. Both the constraints are of the first class, and they unambiguously determine the particle dynamics in $\mathbb{M}_4$ (i.e. the particle world lines).

Since no secondary constraints occur, the system possesses 12 physically essential degrees of freedom. In order to single them out explicitly, two subsidiary *gauge fixing* constraints are needed. They can be given in the general form:

$$\chi(y, r, P, p_\perp, t) = 0, \qquad [\chi, \phi] \neq 0, \quad \partial\chi/\partial t \neq 0. \tag{6.4}$$

$$\psi(y, r, P, p) = 0, \qquad [\psi, r^2] \neq 0. \tag{6.5}$$

These constraints permit one to eliminate redundant time-like variables $x_a^0$ and the corresponding momenta $p_{a0}$, and then to pass to the three-dimensional Hamiltonian description.

The gauge fixing constraints do not influence the dynamics of the model, but their choice determines specific features of the final description, namely, the reduced phase space $\mathbb{P}$ (as a submanifold of $T^*\mathbb{M}_4^2$), the induced Poisson brackets, and a possible choice of variables, in terms of which these brackets take the canonical form. An explicit form of observables (i.e. the covariant particle positions, the generators of the Poincaré group etc.), being functions of the canonical variables of space $\mathbb{P}$, depends on a choice of the gauge fixing constraints, too. Thus, using the arbitrariness of this choice one can make an effective influence on the structure of the final description.

A special choice of the constraint (6.4) in the form

$$\chi = \chi(y, r, P_0, \tau), \tag{6.6}$$

allows one to avoid a well-known *no−interaction theorem* [ 11 ], that is, to pass to such a three-dimensional Hamiltonian description of time-asymmetric models in which the spatial covariant particle positions $x_a^i$ ($a = 1, 2;\ i = 1, 2, 3$) become canonical variables.

The three-dimensional Hamiltonian description in terms of covariant variables is desirable in various aspects. For example, it simplifies the introduction of the interaction with external fields and allows a position representation on the quantum-mechanical level. But this description is not convenient for solving a two-body problem, because it does not provide a relevant separation of external and internal degrees of freedom.

Another choice of the gauge fixing constraint (6.4),

$$\chi = y^0 + \mathrm{tr}(\Lambda\,\Omega\,\partial\Lambda^{\mathrm{T}}/\partial P_0) - \tau = 0, \tag{6.7}$$

where

$$\Omega_{\mu\nu} \equiv r_\mu p_\nu - r_\nu p_\mu, \tag{6.8}$$

$|P| \equiv \sqrt{P^2}$, and matrix $\|\Lambda(P/|P|)^\nu{}_\mu\| \in \mathcal{SO}(1,3)$, $\Lambda^\mu{}_\nu P^\nu = \delta_0^\mu |P|$ describes the Lorentz transformation into the centre-of-mass (CM) reference frame, leads to a three-dimensional Hamiltonian description within the framework of the Bakamjian-Thomas model [ 41, 42, 43 ]. Within this description ten generators of the Poincaré group $P_\mu$, $J_{\mu\nu}$, as well as the covariant particle positions $x_a^\mu$ are the functions of canonical variables $\mathbf{Q}$, $\mathbf{P}$, $\boldsymbol{\rho}$, $\boldsymbol{\pi}$. The only arbitrary function entering into expressions for canonical generators is the total mass $|P| = M(\boldsymbol{\rho}, \boldsymbol{\pi})$ of the system which determines its internal dynamics. For time-asymmetric models this function is defined by the mass-shell equation [ 18, 32 ] which can be derived from the dynamical constraint via the following substitution of arguments on the l.-h.s. of (6.3):

$$P^2 \to M^2, \quad p_\perp^2 \to -\boldsymbol{\pi}^2, \quad P{\cdot}r \to \epsilon M\rho, \quad p_\perp{\cdot}r \to -\boldsymbol{\pi}{\cdot}\boldsymbol{\rho}; \quad \text{here } \rho \equiv |\boldsymbol{\rho}|. \tag{6.9}$$

Due to the Poincaré-invariance of the description, it is sufficient to choose the CM reference frame in which $\mathbf{P} = \mathbf{0}$, $\mathbf{Q} = \mathbf{0}$. Accordingly, $P_0 = M$, $J_{0i} = 0$ ($i = 1, 2, 3$), and the components $S_i \equiv \frac{1}{2}\varepsilon_i{}^{jk} J_{jk}$ form a 3-vector of the total spin of the system (internal angular momentum) which is an integral of motion. At this point the problem is reduced to a rotation-invariant problem of some effective single particle which is integrable in terms of polar coordinates,

$$\boldsymbol{\rho} = \rho\mathbf{e}_\rho, \quad \boldsymbol{\pi} = \pi_\rho\mathbf{e}_\rho + S\mathbf{e}_\varphi/\rho. \tag{6.10}$$

Here $S \equiv |\mathbf{S}|$; the unit vectors $\mathbf{e}_\rho$, $\mathbf{e}_\varphi$ are orthogonal to $\mathbf{S}$, they form together with $\mathbf{S}$ a right-oriented triplet and can be decomposed in terms of the Cartesian unit vectors $\mathbf{i}$, $\mathbf{j}$:

$$\mathbf{e}_\rho = \mathbf{i}\cos\varphi + \mathbf{j}\sin\varphi, \quad \mathbf{e}_\varphi = -\mathbf{i}\sin\varphi + \mathbf{j}\cos\varphi, \tag{6.11}$$

where $\varphi$ is a polar angle.

The corresponding quadratures read:

$$t - t_0 = \int \mathrm{d}\rho \; \partial \pi_\rho(\rho, M, S)/\partial M, \tag{6.12}$$

$$\varphi - \varphi_0 = -\int \mathrm{d}\rho \; \partial \pi_\rho(\rho, M, S)/\partial S, \tag{6.13}$$

where $t$ is an evolution parameter fixed by constraint (6.7) in the CM reference frame, and the radial momentum $\pi_\rho$ as a function of $\rho$, $M$, $S$ is defined by the mass-shell equation written down in terms of these variables,

$$\phi\left(M^2, \; -\boldsymbol{\pi}^2, \; \epsilon M\rho, \; -\boldsymbol{\pi} \cdot \boldsymbol{\rho}\right) \equiv \phi\left(M^2, \; -\pi_\rho^2 - \frac{S^2}{\rho^2}, \; \epsilon M\rho, \; -\pi_\rho\rho\right) = 0. \tag{6.14}$$

The solution of the problem given in terms of canonical variables enables one to obtain particle world lines in the Minkowski space using the following formulae [18, 32]:

$$x_a^0 = t + \frac{1}{2}(-)^{\bar{a}}\epsilon\rho, \tag{6.15}$$

$$\mathbf{x}_a = \frac{1}{2}(-)^{\bar{a}}\boldsymbol{\rho} + \epsilon\rho\frac{\boldsymbol{\pi}}{M} \equiv \left(\frac{1}{2}(-)^{\bar{a}} + \epsilon\frac{\pi_\rho}{M}\right)\rho\mathbf{e}_\rho + \epsilon\frac{S}{M}\mathbf{e}_\varphi. \tag{6.16}$$

Particularly, vector $\mathbf{r} = \mathbf{x}_1 - \mathbf{x}_2 = \boldsymbol{\rho}$ characterizes the relative motion of particles.

# 7. Time-asymmetric models of particle interactions with long-range and confining potentials

The explicit form of $\phi$ (6.3) depends in a complicated manner on the choice of the original Fokker potential. Its construction is the main difficulty which occurs in the analysis of time-asymmetric models. Let us split function $\phi$ into two parts:

$$\phi_f + \phi_{int} = 0, \tag{7.1}$$

where

$$\phi_f = \frac{1}{4}P^2 - \frac{1}{2}(m_1^2 + m_2^2) + (m_1^2 - m_2^2)\frac{p_\perp \cdot r}{P \cdot r} + p_\perp^2 \tag{7.2}$$

corresponds to a free-particle system, and $\phi_{int}$ is to be found. Hereafter we refer to $\phi_{int}$ as the *Hamiltonian potential*.

Only few cases are known when function $\phi_{int}$ can be constructed explicitly. They correspond to the three-parametric Fokker potential

$$U = U_s + U_v + U_c = (\alpha_s + \alpha_v\omega + \alpha_c\sigma_1\sigma_2)G_\epsilon(r) \;, \tag{7.3}$$

where $\alpha_s, \alpha_v, \alpha_c$ are arbitrary constants. The first and the second terms on the r.-h.s. of equation (7.3) correspond to the scalar and vector field-type interactions with the coupling constants $\alpha_s$ and $\alpha_v$, respectively, and the third term describes the confinement interaction (when $\alpha_c > 0$). In the non-relativistic limit this model leads to the potential $U^{(0)} = (\alpha_s + \alpha_v)/r + \alpha_c r$, where $r$ is the distance between the particles.

The corresponding Hamiltonian potential has the form:

$$
\begin{aligned}
\phi_{int} = & -\frac{2\alpha_s m_1 m_2 + \alpha_v(P^2 - m_1^2 - m_2^2)}{\epsilon P \cdot r} - 2\alpha_c \left( \frac{b_1 b_2}{\epsilon P \cdot r} - \alpha_v \right) \\
& - (\alpha_s^2 - \alpha_v^2) \frac{2\alpha_s m_1 m_2 + (b_1 - \alpha_v)m_2^2 + (b_2 - \alpha_v)m_1^2}{\epsilon P \cdot r((b_1 - \alpha_v)(b_2 - \alpha_v) - \alpha_s^2)}
\end{aligned}
\tag{7.4}
$$

where

$$
b_a \equiv \epsilon(\tfrac{1}{2}P \cdot r + (-)^{\bar{a}} p_\perp \cdot r). \tag{7.5}
$$

It is worth noting that the interactions are combined in terms of the Hamiltonian potential in a non-linear manner.

For other Fokker potentials approximation methods (such as coupling constant expansion) should be applied for the Hamiltonization procedure. Especially for the time-asymmetric analogue of the Fokker potential (5.12) the Hamiltonian potential in the second order approximation in coupling constant $\alpha = g_1 g_2$ reads:

$$
\phi_{int} = -\frac{2m_1 m_2}{\epsilon P \cdot r} \alpha f(\nu) - \frac{\alpha^2 h(\nu)}{\epsilon P \cdot r} \left( \frac{m_1^2}{b_1} + \frac{m_2^2}{b_2} \right) + O(\alpha^3), \tag{7.6}
$$

where

$$
h(\nu) \equiv ((f(\nu) - \lambda f'(\nu))^2 - (f'(\nu))^2 \tag{7.7}
$$

and

$$
\nu \equiv \frac{P^2 - m_1^2 - m_2^2}{2m_1 m_2}. \tag{7.8}
$$

We note that particular cases of (5.12) are the Fokker potentials which correspond to the particle interaction via massless linear tensor fields of an arbitrary rank (see equation (5.13)) and their superpositions.

Below we consider some most interesting features of time-asymmetric models described in this section.

## 7.1. Vector and scalar models

We begin with vector and scalar time-asymmetric interactions. These models are based on the Fokker-type integrals (5.1) with the Fokker potentials $U = U_v$ and $U = U_s$, respectively (see (7.3)). Both scalar and vector models were partly considered earlier (the former in the two-dimensional space-time only) [ 44, 45, 46, 26, 20, 28, 47]. Our results obtained by means of both the Lagrangian and especially the Hamiltonian formulations of these models complete the analysis of their classical dynamics.

The vector and scalar time-asymmetric models present two-body problems lying near the border line of those problems the solution of which can be presented in a closed form. A lot of analyses can be made analytically. Especially, turning and other important for the integration points are solutions of the third and fourth order algebraic equations, while the quadratures cannot be expressed even in terms of elliptic and other special functions and, thus, they need computer work. For simplicity here we limit ourselves to the case of equal particle rest masses $m_0$.

In the non-relativistic limit the vector and scalar interactions reduce to the Coulomb interaction with the coupling constant $\alpha$ (namely, $\alpha_v$ and $\alpha_s$, respectively). Thus, it is convenient to present the specific features of vector and scalar models in comparison with the non-relativistic Coulomb system.

The variety of solutions to the equations of motion of a two-particle system consists of a 12–parametric family. The Poincaré transformations (which form a 10–parametric group) change only the motion of system considered as the whole. This motion does not reflect specific features of the models. Here we do not distinguish solutions which differ from one another by the Poincaré transformations. So, non-equivalent solutions form a two–parametric family. It can be parametrized by values of total mass $M$ (or energy $E$ in the non-relativistic case) and spin (internal angular momentum) $S$, the pair of integrals of motion. Thus, a variety of all the possible solutions is reduced to some subset of $(M, S)$–plane. We note that parameters $m_0$ and $|\alpha|$ become unessential when using $m_0$ and $r_0 \equiv |\alpha| /m_0$ as units of measurement for momentum– and position–like variables, respectively. We also introduce dimensionless integrals of motion $\mu = \frac{1}{2}M/m_0$ and $\sigma = S/ |\alpha|$. For the convenience we will speak about various solutions (namely, phase trajectories, particle trajectories and world lines) as if each of them is placed at the corresponding point of $(\mu, \sigma)$–plane.

First of all we shall consider a vector model. Qualitatively different types of the phase trajectory (three top graphs of figure 1) correspond to three different domains $\mathcal{D}(+)$, $\mathcal{D}(-)$ and $\mathcal{D}(0)$ of $(\mu, \sigma)$–plane (the bottom graph of figure 1).

The number and the position of $(\mu, \sigma)$–domains on $(\mu, \sigma)$–plane are roughly in accordance with the non-relativistic case, while the phase trajectories are more complicated: they consist of few disconnected branches. It does means that there exist few solutions of the Hamiltonian equations of motion at the same values of the integrals of motion $\mu, \sigma$.

Only one branch, namely, $\gamma_-^a$ for the attraction case ($\alpha < 0$) and $\gamma_-^r$ for the repulsion case ($\alpha > 0$), is *regular, i.e.* it is a relativistic analogue of the phase trajectory of the Coulomb system and coincides with the latter in a weakly relativistic domain of $(\mu, \sigma)$–plane (*i.e.*, $\mu \approx 1$, and $\sigma \gg 1$ for the attraction case). If $\mu > 1$ ($\mathcal{D}(+)$ domain of $(\mu, \sigma)$–plane), both $\gamma_-^a$ and $\gamma_-^r$ exist and correspond to unbounded particle trajectories which are analogues to the hyperbolas of the Coulomb problem. We note that particle trajectories $\gamma_-^a$ have a loop-like shape at the points of $(\mu, \sigma)$–plane which are close to $\mu = 1$, $\sigma = 1$ (figure 2a). This effect becomes more evident for bounded states $\gamma_-^a$ (existing in $\mathcal{D}(-)$) as the appearance of the perihelion advance (figure 2b) (states $\gamma_-^r$ of the repulsion case disappear in this domain). In the ultrarelativistic case $\mu \to 0$ (the left lower corner of $\mathcal{D}(-)$) the particles "stick" together, so that the distance between them becomes far less than the distance to the centre-of-mass (figure 2c). On curve $\mathcal{F}$ the regular states correspond to a circular motion of particles, so that in domain $\mathcal{D}(0)$ (from below curve $\mathcal{F}$) the regular

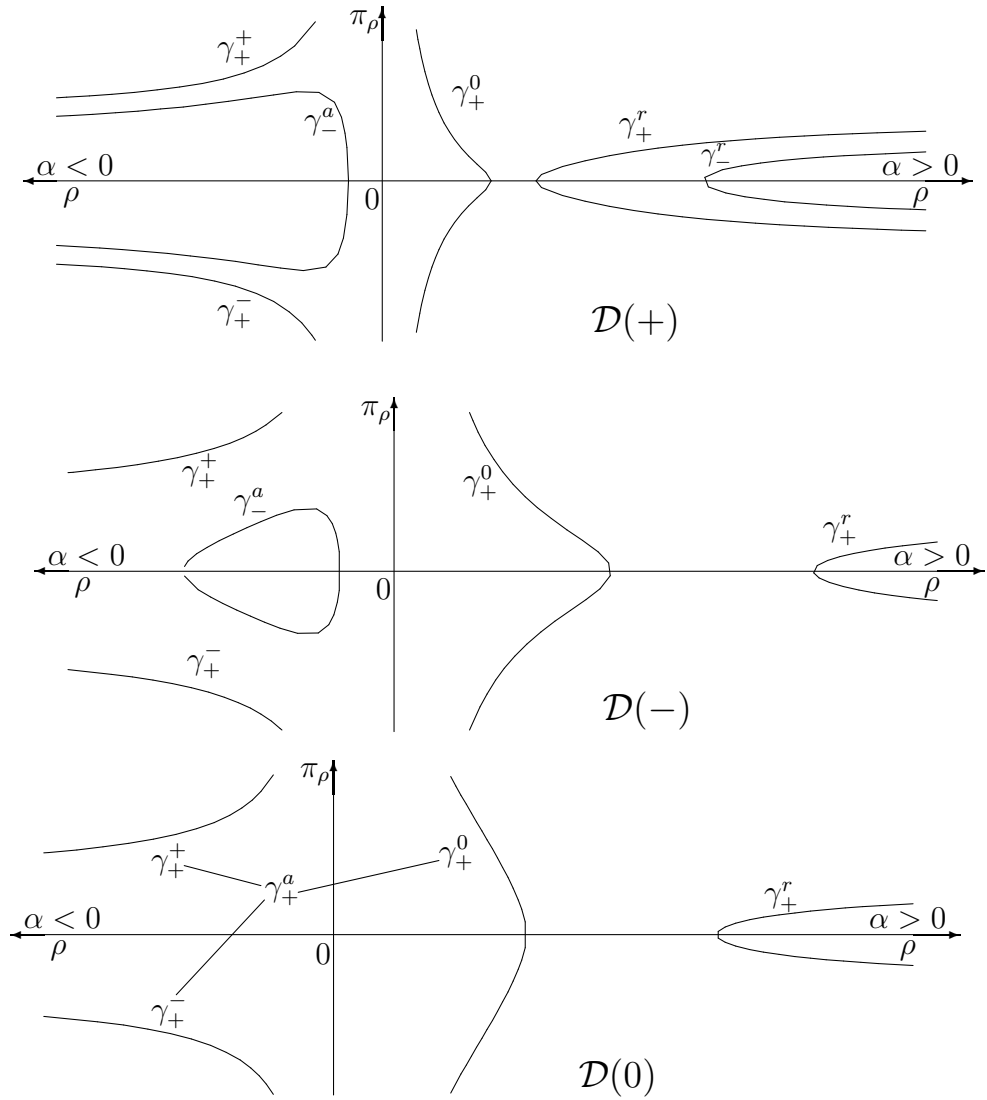

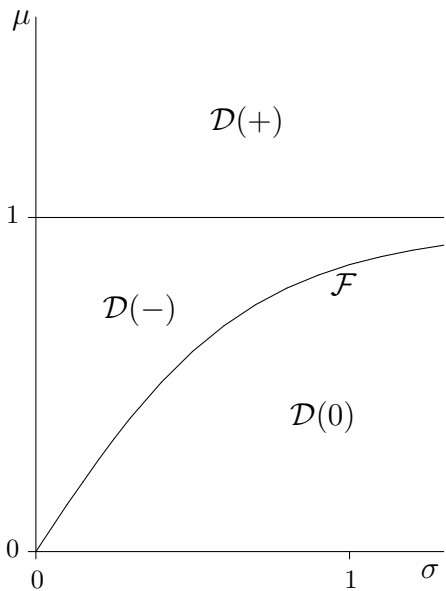

Figure 1. Vector model. Various domains of $(\mu, \sigma)$-plane (bottom graph) and the corresponding types of the phase trajectory (three top graphs). Curve $\mathcal{F}$ on the $(\mu, \sigma)$-plane is determined by the parametric equations:

$$\sigma^2 = \frac{p(p+2)^2}{2p^2+5p+4},$$

$$\mu^2 = \frac{p(2p^2+5p+4)}{2(p+4)^2},$$

$$p \in [0, \infty[.$$

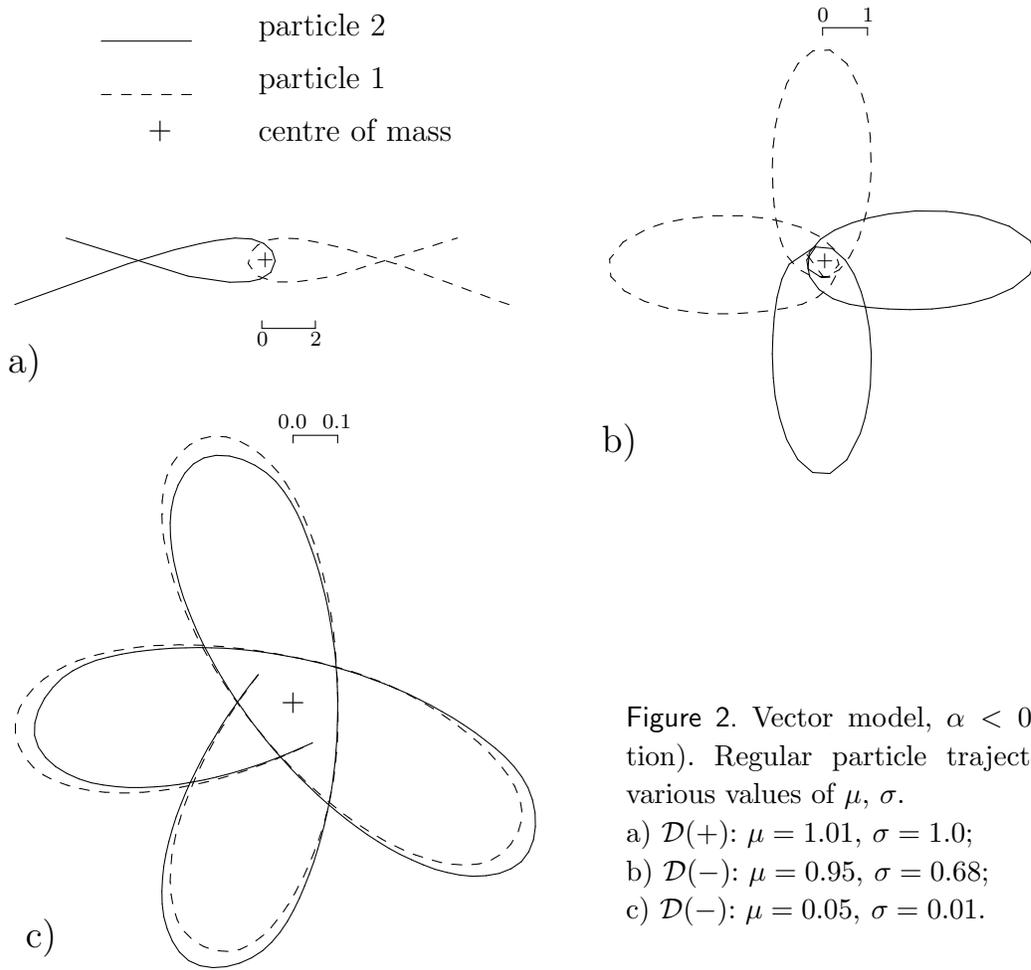

Figure 2. Vector model, $\alpha < 0$ (attraction). Regular particle trajectories for various values of $\mu$, $\sigma$.
a) $\mathcal{D}(+)$: $\mu = 1.01$, $\sigma = 1.0$;
b) $\mathcal{D}(-)$: $\mu = 0.95$, $\sigma = 0.68$;
c) $\mathcal{D}(-)$: $\mu = 0.05$, $\sigma = 0.01$.

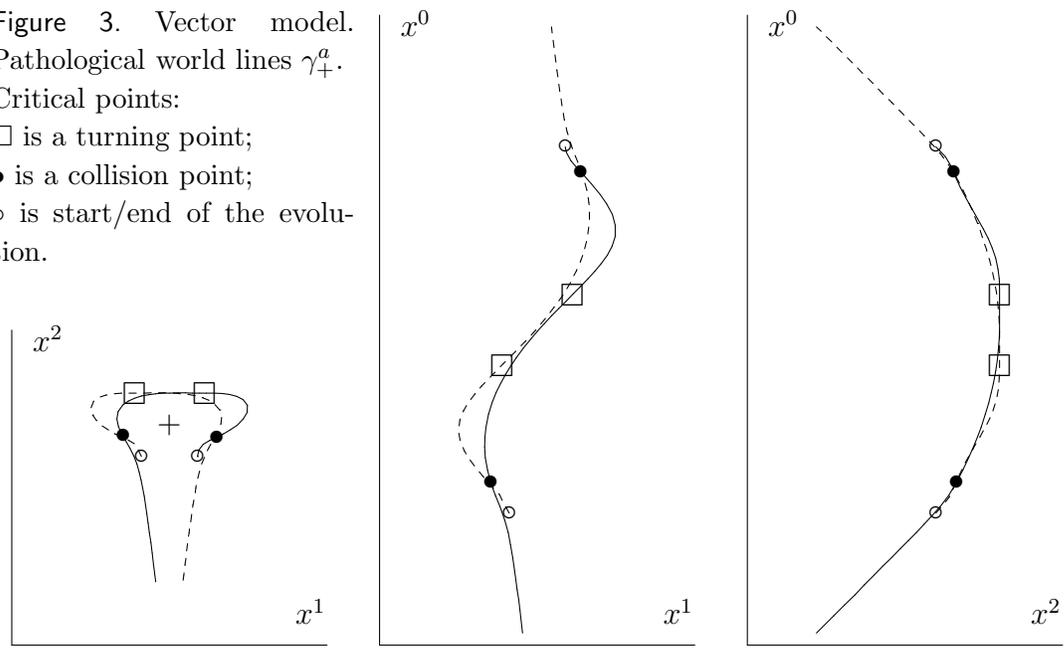

Figure 3. Vector model. Pathological world lines $\gamma_+^a$. Critical points:
□ is a turning point;
● is a collision point;
○ is start/end of the evolution.

motion is forbidden.

The remaining branches $\gamma_+^0$, $\gamma_+^+$, $\gamma_+^-$, and $\gamma_+^r$ of the phase trajectory do not have non-relativistic counterparts. They exist and are qualitatively similar on the whole $(\mu, \sigma)$–plane. These branches present a rather strange motion of particles, so that the sign of $\alpha$ does not characterize the interaction as attractive or repulsive. Moreover, it turns out natural to sew up the three branches $\gamma_+^0$, $\gamma_+^+$, and $\gamma_+^-$ into a unique one $\gamma_+^a$ (this is shown in figure 1 for the phase trajectory in $\mathcal{D}(0)$), so that the resulting motion is as follows: the particles move from an infinite distance between them to their collision, go through one another and go away to the distance $\sim r_0$, draw closer to one another, collide again, and go away to an infinite distance (figure 3). Branches $\gamma_+^r$ and $\gamma_+^a$ and the corresponding world lines are *pathological* in the sense that the velocities of massive particles tend asymptotically up to the light speed. Besides, these solutions contain critical points (namely, collision and turning points) in which massive particles reach but not exceed the light speed at a finite time. Nevertheless, the particle world lines turn out smooth both at these points and everywhere. Another specific feature of the pathological states is that the evolution of particles is spread over a semiinfinite interval of the coordinate time while the evolution parameter covers the whole real axes (figure 3).

The scalar model is more intricate, especially for an attractive interaction. There are more qualitatively different types of the phase trajectory which correspond to a larger number of $(\mu, \sigma)$–domains and which consist of more branches (figure 4).

Among them only one branch is regular, *i.e.* analogous to the Coulomb phase trajectory. It exists in the domains $\mathcal{D}(1\pm; 1)$. Bounded states (in $\mathcal{D}(1-; 1)$) present a motion of particles with the perihelion retardance (unlike the advance in the vector model). They disappear from below curve $\mathcal{F}$, $\sigma > 1/\sqrt{5}$, on which the particle trajectories become circular.

In contrast to the case of a vector model, the domain of regular states is bounded not only from below, but also from the left where a motion is not forbidden. The border lines $\mathcal{X}_+$ and $\mathcal{J}$, $\mu > \sqrt{5/8}$ indicate no special changes in the particle motion except the appearance of critical points (which corresponds to reaching the light speed) on the particle world lines. These *provisionally regular* states exist in the domains $\mathcal{D}(1\pm; 2)$ and $\mathcal{D}(1\pm; 3)$. The effect of the perihelion retardance grows for them (figure 5a), especially in the domain $\mathcal{D}(1\pm; 3)$; here the particles move as if they attract one another at a large distance, while at a small distance $\sim r_0$ each particle repulses another one by a very (but not absolutely) hard core. The particles bounce back off this core with the light speed, but their world lines are smooth at this critical point (figure 5b). Going to curve $\mathcal{J}$, $\sigma < 1/\sqrt{5}$, the particle trajectories tend (as in the regular case) to circular ones, but in a very strange manner: the particles rebound more frequently (figure 5c), so that in the limiting circular trajectories (which corresponds to $\mathcal{J}$ itself) the set of critical points becomes dense everywhere.

Apart from the regular or provisionally regular states (which present a reasonable behaviour of the particles on the whole) and the pathological ones (which are roughly similar to those in the vector model), the attraction (i.e. $\alpha < 0$) scalar model possesses some *exotic* states which correspond to a bounded particle motion

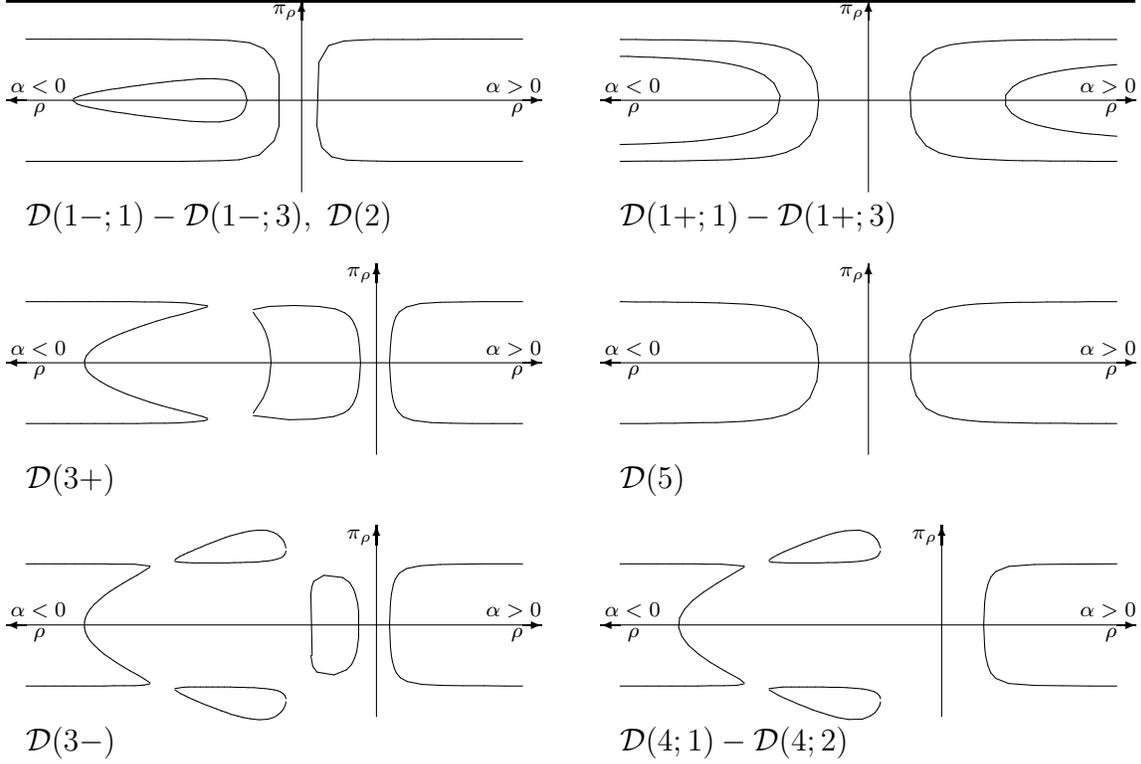

$\mathcal{D}(1-;1) - \mathcal{D}(1-;3),\ \mathcal{D}(2)$      $\mathcal{D}(1+;1) - \mathcal{D}(1+;3)$

$\mathcal{D}(3+)$      $\mathcal{D}(5)$

$\mathcal{D}(3-)$      $\mathcal{D}(4;1) - \mathcal{D}(4;2)$

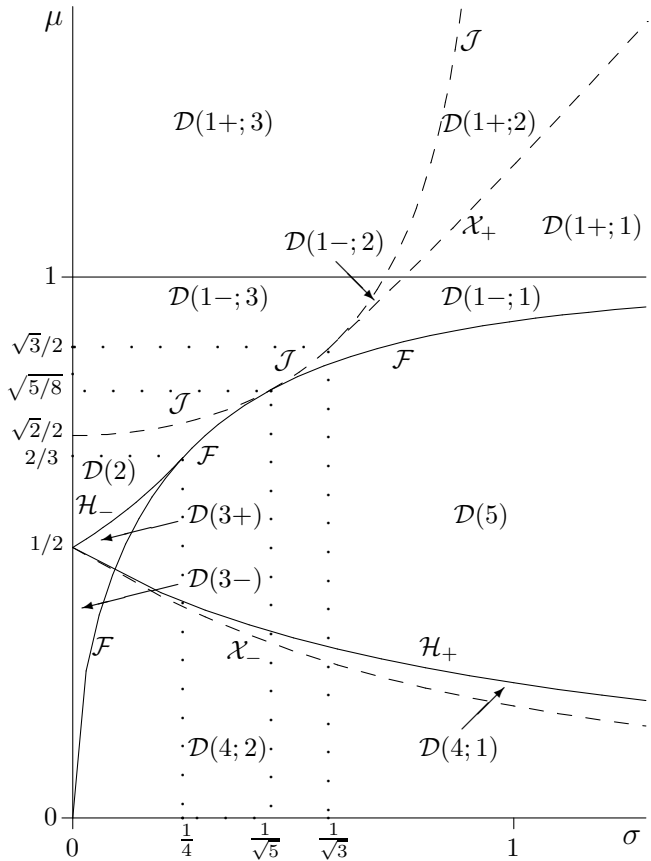

**Figure 4.** Scalar model. Various domains of the $(\mu, \sigma)$-plane (bottom graph) and the corresponding types of the phase trajectory (six top graphs).

The curves $\mathcal{F}$, $\mathcal{J}$, $\mathcal{H}_{\pm}$, and $\mathcal{X}_{\pm}$ on the $(\mu, \sigma)$-plane are defined by the equations:

$$\mathcal{F}: \quad \mu^2 = \frac{27\sigma}{2\left((3+\sigma^2)^{3/2} + \sigma(9-\sigma^2)\right)},$$

$$\mathcal{J}: \quad \mu^2 = \frac{1}{2(1-\sigma^2)},$$

$$\mathcal{H}_{\pm}: \quad \mu = \frac{1}{2(1\pm\sigma)},$$

$$\mathcal{X}_{\pm}: \quad \mu = \left(\sqrt{1+\sigma^2} \pm \sigma\right)/2.$$

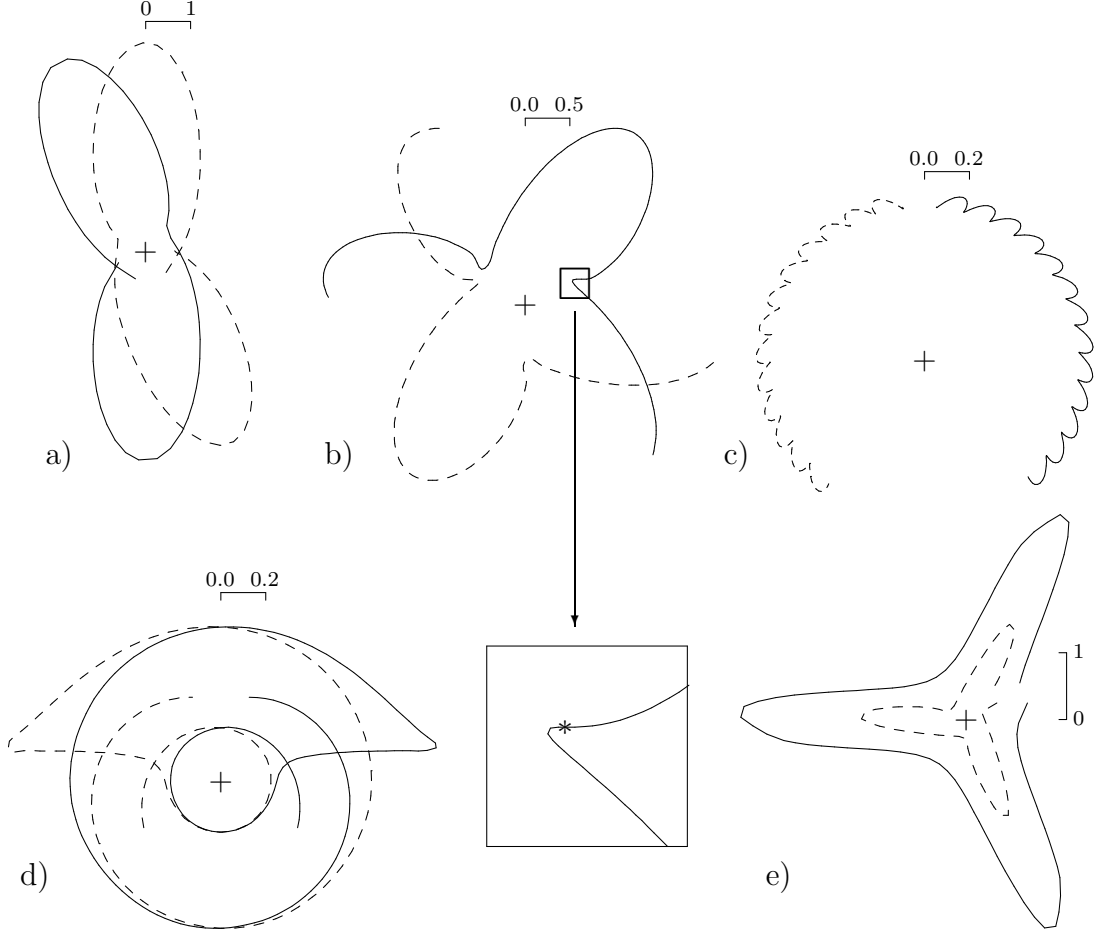

**Figure 5.** Scalar model, $\alpha < 0$ (attraction). Various types of bounded particle trajectories.
a) $\mathcal{D}(1-;2)$ : $\mu = 0.95$, $\sigma = 0.68$; b) $\mathcal{D}(1-;3)$ : $\mu = 0.9$, $\sigma = 0.5$;
c) $\mathcal{D}(1-;3)$ near $\mathcal{J}$ : $\mu = 0.75$, $\sigma = 0.3$; d) $\mathcal{D}(2)$ near $\mathcal{H}_-$ : $\mu = 0.6$, $\sigma = 0.16666$;
e) $\mathcal{D}(4;2)$ : $\mu = 0.3$, $\sigma = 0.15$.

at a relative distance of order $r_0$. These states exist in the domains $\mathcal{D}(2)$, $\mathcal{D}(3\pm)$, and $\mathcal{D}(4;1) - \mathcal{D}(4;2)$, i.e., far from the weakly relativistic domain, and thus they have no non-relativistic analogues. For example, in the domain $\mathcal{D}(2)$ the particles move as if each particle repulses another one by the hard exterior of an empty core inside (figure 5d); in the domains $\mathcal{D}(4;1) - \mathcal{D}(4;2)$ the trajectory of one of the particles always lies inside the trajectory of another particle (figure 5e).

The variety of solutions described above is obtained within the Hamiltonian formulation of the vector and scalar models. Within the framework of the Lagrangian formalism only regular solutions can be reconstructed completely. Besides, this framework partially recovers provisionally regular solutions, namely, some segments of world lines between the critical points. Other solutions disappear within the Lagrangian formalism.

In our consideration the Lagrangian formalism is primary with respect to the Hamiltonian one. Thus, one can conclude at first sight that non-Lagrangian solutions have no physical meaning. On the other hand, the Hamiltonian formulation of the models is an important link toward their quantization, and non-Lagrangian solutions may contribute to the resulting quantum-mechanical picture.

These complicated questions are discussed in more detail in sections 8 and 9.1 where we study the classical and quantum mechanics of the vector and scalar models in $\mathbb{M}_2$.

### 7.2. Scalar–vector model

The purely vector and scalar time-asymmetric models are calculatingly cumbersome and present a rather intricate particle dynamics. The case of arbitrary superposition of the scalar and vector interaction is not expected to be simpler (though it is also solvable). It follows from the complicated structure of the Hamiltonian potential (see equation (7.4) with $\alpha_c = 0$).

In a special case of superposition, $\alpha_v = \kappa\alpha_s \equiv \kappa\alpha$, $\kappa = \pm 1$, the second term of $\phi_{int}$ (7.4) vanishes. This structure of the dynamical constraint simplifies to a great extent the dynamics of the model and makes it similar, in the mathematical respect, to the dynamics of a non-relativistic system with the Coulomb interaction. In this case one can expect the existence of an additional integral of motion, the relativistic analogue of the Runge-Lenz vector.

Actually, it is easy to guess the structure of this integral of motion working within the framework of manifestly covariant Hamiltonian mechanics [31]. For this purpose it is convenient to simplify the free-particle term $\phi_f$ (10) of the dynamical constraint whose cumbersome form obscures the following treatment of the model and is caused by a descriptional rather than dynamical reason. Let us perform the canonical transformation $(y^\mu, \; P_\mu, \; r^\mu, \; p_\mu) \longmapsto (z^\mu, \; P_\mu, \; r^\mu, \; q_\mu)$,

$$q_\mu = p_\mu - \frac{m_1^2 - m_2^2}{2P^2}P_\mu, \qquad z^\mu = y^\mu + \frac{m_1^2 - m_2^2}{2P^2}\Big(r^\mu - 2\frac{P \cdot r}{P^2}P^\mu\Big), \qquad (7.9)$$

(the variables $r^\mu$ and $P_\mu$ remain unchanged). In terms of new variables the dynamical constraint takes the form:

$$\phi = \frac{1}{4}P^2 - \frac{1}{2}(m_1^2 + m_2^2) + \frac{(m_1^2 - m_2^2)^2}{4P^2} + q_\perp^2 - \frac{\alpha(P^2 - (m_1 - \kappa m_2)^2)}{\epsilon P \cdot r} = 0, \quad (7.10)$$

where

$$q_{\perp\mu} \equiv P^\nu \Xi_{\nu\mu}/P \cdot r, \quad \Xi_{\mu\nu} = r_\mu q_\nu - r_\nu q_\mu; \quad q_\perp \cdot P \equiv 0. \qquad (7.11)$$

Then, it is easy to examine that the relativistic analogue of the Runge-Lenz vector has the following form:

$$R_\mu = \Pi_\mu^\nu\Big(q_\perp^\lambda \Xi_{\lambda\nu} + \frac{\alpha(P^2 - (m_1 - \kappa m_2)^2)}{2\epsilon P \cdot r}r_\nu\Big), \qquad (7.12)$$

where $\Pi_\mu^\nu \equiv \delta_\mu^\nu - P_\mu P^\nu/P^2$. It is indeed an integral of motion, i.e.

$$[R_\mu, \phi] \approx 0, \qquad [R_\mu, r^2] = 0 \qquad (7.13)$$

and satisfies the relations:

$$[R_\mu, P_\nu] = 0, \qquad [R_\mu, J_{\lambda\sigma}] = -\eta_{\mu\lambda} R_\sigma + \eta_{\mu\sigma} R_\lambda, \tag{7.14}$$

$$[R_\mu, R_\nu] \approx \left( \frac{1}{4} P^2 - \frac{1}{2}(m_1^2 + m_2^2) + \frac{(m_1^2 - m_2^2)^2}{4P^2} \right) \Pi_\mu^\lambda \Pi_\nu^\sigma J_{\lambda\sigma}, \tag{7.15}$$

where the Dirac symbol $\approx$ denotes a weak equality.

The relations (7.14)–(7.15) are similar to those obtained for the Runge–Lenz vector of a simple relativistic oscillator and Coulomb models in [ 48 ]. These relations are essentially nonlinear and thus their group theoretical treatment is complicated. In the present paper we limit our study to the case of the CM reference frame in which the corresponding Poisson bracket relation can be linearized.

For this purpose we reformulate (as in the previous cases) the present time-asymmetrical model into the framework of the Bakamjian-Thomas model. Then the Runge-Lenz vector becomes $R_\mu = (0, \mathbf{R})$, where

$$\mathbf{R} = \boldsymbol{\pi} \times \mathbf{S} + g(M) \boldsymbol{\rho}/\rho, \tag{7.16}$$

$\mathbf{S} = \boldsymbol{\rho} \times \boldsymbol{\pi}$ is a spin of the system, and the total mass satisfies the equation

$$d(M) - \boldsymbol{\pi}^2 - 2g(M)/\rho = 0. \tag{7.17}$$

Here

$$d(M) \equiv \frac{1}{4M^2} \left( M^2 - (m_1 + m_2)^2 \right) \left( M^2 - (m_1 - m_2)^2 \right), \tag{7.18}$$

$$g(M) \equiv \frac{\alpha}{2M} \left( M^2 - (m_1 - \kappa m_2)^2 \right). \tag{7.19}$$

Besides, in the CM reference frame the covariant particle positions are the following functions of the canonical variables:

$$\mathbf{x}_a = \frac{(-)^{\bar{a}}}{2} \left( 1 + \frac{m_{\bar{a}}^2 - m_a^2}{M^2} \right) \boldsymbol{\rho} + \epsilon \rho \frac{\boldsymbol{\pi}}{M}, \qquad a = 1, 2; \quad \bar{a} \equiv 3 - a. \tag{7.20}$$

The Poisson bracket relations for the internal angular momentum (spin) of the system $\mathbf{S}$ and the Runge-Lenz vector $\mathbf{R}$ are similar to those in the non-relativistic Coulomb problem:

$$\{S_i, S_j\} = \varepsilon_{ij}{}^k S_k, \quad \{R_i, S_j\} = \varepsilon_{ij}{}^k R_k, \quad \{R_i, R_j\} = -d(M) \varepsilon_{ij}{}^k S_k. \tag{7.21}$$

Indeed, when $d(M) = 0$, equations (7.21) are the relations for generators of the Euclidian group $\mathcal{E}(3)$. In the case $d(M) \neq 0$ the $S_i$ and the normalized $\hat{R}_i \equiv R_i / \sqrt{|d|}$ generate the group $\mathcal{SO}(4)$, when $d(M) < 0$, and the group $\mathcal{SO}(1,3)$, when $d(M) > 0$. Taking into account equation (7.21) we obtain the following cases for the algebra of internal symmetries:

$$\begin{aligned} &\mathfrak{so}(4) && \text{for } |m_1 - m_2| < M < m_1 + m_2, \\ &\mathfrak{e}(3) && \text{for } M = |m_1 - m_2| \text{ and } M = m_1 + m_2, \end{aligned}$$

$\mathfrak{so}(1,3)$  for $0 < M < |m_1 - m_2|$ and $M > m_1 + m_2$.

The existence of the Runge–Lenz vector makes it possible to obtain both the relative and particle trajectories traced by vectors $\boldsymbol{\rho}$ and $\mathbf{x}_a$, respectively, without an integration. At first we note that these trajectories are flat curves placed on the plane orthogonal to the spin of the system, i.e. $\boldsymbol{\rho} \cdot \mathbf{S} = \mathbf{x}_a \cdot \mathbf{S} = 0$. Vector $\mathbf{R}$ lies on the same plane, i.e. $\mathbf{R} \cdot \mathbf{S} = 0$. Multiplying equation (7.16) by $\boldsymbol{\rho}$ one can obtain the relation:

$$\mathbf{R} \cdot \boldsymbol{\rho} = g\rho + S^2, \tag{7.22}$$

where $S \equiv |\mathbf{S}|$. Let $\varphi$ be an angle between $\mathbf{R}$ and $\boldsymbol{\rho}$, i.e. $\mathbf{R} \cdot \boldsymbol{\rho} = R\rho \cos \varphi$. Then equation (7.22) can be reduced to the canonical equation of a conic section

$$p/\rho = e \cos \varphi - \operatorname{sgn} g \tag{7.23}$$

with the following canonical parameter $p$ and eccentricity $e$:

$$p = \frac{S^2}{|g|} = \frac{2MS^2}{|\alpha||M^2 - (m_1 - \kappa m_2)^2|}, \quad e = \frac{R}{|g|} = \sqrt{1 + \frac{S^2}{\alpha^2}\frac{M^2 - (m_1 + \kappa m_2)^2}{M^2 - (m_1 - \kappa m_2)^2}}. \tag{7.24}$$

Searching for the equations of particle trajectories is a similar but somewhat complicated task. Let us define the vectors:

$$\mathbf{r}_a \equiv \mathbf{x}_a - c_a \mathbf{R}, \tag{7.25}$$

where

$$c_a = \frac{2(-)^{\bar{a}}}{(M + m_{\bar{a}})^2 - m_a^2}. \tag{7.26}$$

Then, one can obtain the relations

$$(-)^{\bar{a}}\mathbf{R} \cdot \mathbf{r}_a = g r_a + \frac{m_{\bar{a}}}{M}S^2, \tag{7.27}$$

which are similar to equation (7.22) and hence can be written down as follows:

$$p_a/r_a = e \cos \varphi_a - \operatorname{sgn} g, \tag{7.28}$$

where $\varphi_a$ are angles between $(-)^{\bar{a}}\mathbf{R}$ and $\mathbf{r}_a$. Equations (7.28) describe the particle trajectories as being conic sections of the same shape as the relative trajectory, i.e. with the same eccentricity $e$ (7.25) but with other canonical parameters $p_a = \frac{m_{\bar{a}}}{M}p$. The foci of these conic sections are shifted with respect to the centre of mass by vectors $c_a\mathbf{R}$. On the contrary, the non-relativistic particle trajectories have a common focus which is located in the centre of mass.

### 7.3. Models with higher rank tensor interactions.

As it was pointed out above, among time-asymmetric field-type models only those corresponding to the (arbitrary) superposition of scalar and vector interactions permit the exact hamiltonization. In the case when the rank of the field $n \geq 2$, the transition to the Hamiltonian description and the construction of quadratures can be done by means of the method of expansion in a coupling constant.

The structure of the second order Fokker potential (7.6) is common for linear field-type interactions of various tensor dimensions. It specifies the sort of interaction by the functions $f(\nu)$ and $h(\nu)$ which depend on the integral of motion $\nu$ only. Moreover, the nonlinear gravitational interaction can be also described (at least in a slow motion approximation) by this potential (7.6) (see [49, 50]) with

$$f_{gr}(\nu) = 2\nu^2 - 1, \tag{7.29}$$

$$h_{gr}(\nu) = -2(2\nu^2 + 1), \tag{7.30}$$

and $\alpha_{gr} = -\Upsilon m_1 m_2$ where $\Upsilon$ is the gravitational constant. It is possible to integrate a two-body problem considering $f$ and $h$ as arbitrary first and second order functions, respectively.

We note that in the second order approximation the quadratures for the present case can be expressed in terms of elementary functions. For bounded states they lead to the relative motion trajectory of a very simple form,

$$1/\rho = |a| + b\cos((1-\delta)\varphi) \qquad (b < |a|), \tag{7.31}$$

where $a$, $b$, and $\delta$ are functions of the integrals of motion. It describes an ellipse which precesses with the perihelion advance

$$\Delta\varphi = 2\pi\delta = -\pi\alpha^2 h(1)/S^2. \tag{7.32}$$

In the case of a linear purely tensor interaction of arbitrary rank $n$ the perihelion advance $\Delta\varphi$ can be calculated by means of the formulae (7.7), (5.12)–(5.13),

$$\Delta\varphi = \pi(2n^2 - 1)(g_1 g_2/S)^2. \tag{7.33}$$

For the gravitational interaction, using (7.30), we obtain

$$\Delta\varphi = 6\pi(\Upsilon m_1 m_2/S)^2. \tag{7.34}$$

The spatial particle trajectories calculated by means of (7.16) turn out to be more intricate than the relative trajectory which is the typical feature of time-asymmetric models. Nevertheless, their analysis leads to the same value of the perihelion advance.

We note that these relations for the perihelion advance fit those obtained within the various quasirelativistic approaches to the relativistic direct interactions [51, 52, 53, 1, 54].

## 7.4. Confinement models

Our simplest version of a confinement model [ 55] is based on the Fokker potential $U_c$ (see equation (7.3)), the time-asymmetric counterpart of which is proposed in [ 37]. This model could be regarded as a classical relativisation of the primitive quarkonium model with the linear non-relativistic potential. Of course, the relativisation of any non-relativistic system is not unique. There exists in the literature a wide variety of relativistic versions of the potential confinement model. The present model has a number of features which are expected for the models of this kind but which usually are not realized together.

1. The model is a self-consistent relativistic two-particle model. The quantities in terms of which it is built have a clear physical meaning. Solutions of this model are free of any critical point and lead to timelike particle world lines.

2. It is well known that a non-relativistic potential model with the linear potential leads to the Regge trajectory with the unsatisfactory asymptote $M \sim S^{2/3}$. Here we do not propose a quantum version of the present model, but we make the estimates of the Regge trajectory from what follows.

Usually the Regge trajectories in the potential models are calculated in the oscillator approximation [ 56]. Then, the leading Regge trajectory originates from the classical mechanics: it coincides with the curve of circular motions on the $(M, S)$–plane. In our case this curve is described by the following equation:

$$S = \frac{M^2(1 - 4m_0^2/M^2)^{3/2}}{6\sqrt{3}\alpha_c} \qquad (7.35)$$

(we consider the case of equal particle rest masses $m_0$). In the ultrarelativistic limit $M \to \infty$ it leads to the desirable linear asymptote:

$$M^2 \approx 6\sqrt{3}\alpha_c S. \qquad (7.36)$$

It is remarkable that this asymptote is achieved only by taking account of relativity.

3. The present model permits the interpretation of an interaction in terms of some classical fields. It follows from the fact that the Fokker potential $U_c$ can be transformed into an equivalent form,

$$\tilde{U}_c = -2\alpha_c \omega D_\epsilon(x), \qquad (7.37)$$

where function $D_\epsilon(x)$,

$$D_\epsilon(x) = \tfrac{1}{2}\Theta(\epsilon x^0)\Theta(x^2), \qquad (7.38)$$

is the fundamental solution of the equation

$$\Box^2 D_\epsilon(x) = 4\pi\delta(x). \qquad (7.39)$$

Thus, the interaction of particles can be considered as mediated by the vector field obeying some fourth order equation. Gauge invariant nonlinear equations of this kind arise when considering the behaviour of a gluon propagator in the infrared region [57]. Static solutions of such equations are used in a sort of the bag model of confinement [58].

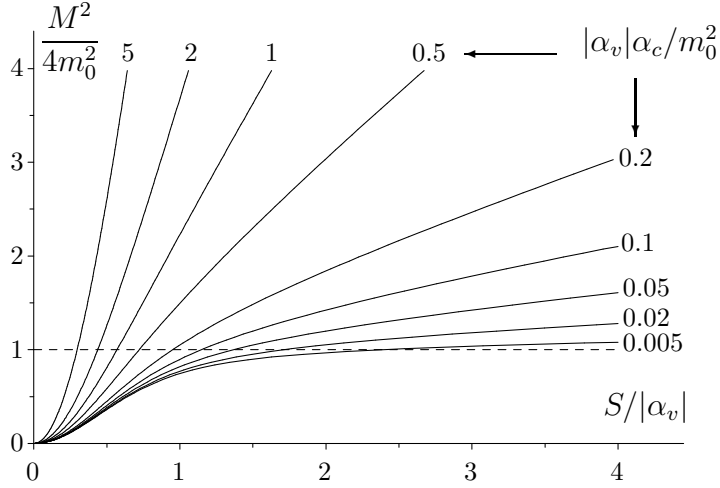

**Figure 6.** Confinement model. Classical Regge trajectories at various rates of the coupling constants and the rest mass.

The simplest version of the relativistic confinement model can be appropriate for the description of light mesons for which the confinement interaction dominates. To include into consideration also heavy mesons one can modify the present model by adding to $U_c$ the usual vector potential $U_v$ (with the appropriate coupling constant $\alpha_v < 0$) [55]. In the non-relativistic limit this mixture leads to the well known potential $U^{(0)} = -|\alpha_v|/r + \alpha_c r$. The resulting model becomes appreciably cumbersome but still remains solvable. Pathological solutions which occur in this model can be unambiguously separated from its regular solutions (which are free of critical points). As an illustration we present the classical Regge trajectories for various rates of the coupling constants and the rest mass (figure 6). We note that all the trajectories tend asymptotically to straight lines. Moreover, the vector correction does not influence their asymptotic behaviour which is still described by equation (7.36).

## 8. Vector and scalar models in $\mathbb{M}_2$

The analysis of the vector and scalar time-asymmetric models in the four-dimensional space-time $\mathbb{M}_4$ was carried out in the previous section for the case of equal particle masses. The cumbersome form of the expressions and a large set of possible motions obscures the physical understanding of the obtained results. In the two-dimensional space-time $\mathbb{M}_2$ the analysis of dynamics becomes considerably simpler even for different particle masses. The dynamics in $\mathbb{M}_2$ seems to correspond to the motions with the inner angular momentum (spin) $S = 0$. But as it turns out, the limit $S \to 0$ is a singular one. Therefore, the consideration of the dynamics of such models in the two-dimensional Minkowski space $\mathbb{M}_2$ appears to be interesting.

The Fokker-type action integral with a time-asymmetric variant of the Fokker potential (5.12) in the front form in $\mathbb{M}_2$ leads to the Lagrangian [ 17]

$$L = -\sum_{a=1}^{N} m_a k_a - \frac{\alpha k_1 k_2 f(\omega)}{r}, \qquad r > 0, \tag{8.1}$$

where

$$\omega = \frac{1}{2}\left(\frac{k_1}{k_2} + \frac{k_2}{k_1}\right). \tag{8.2}$$

The existence of three integrals of motion, which for the Lagrangian (8.1) have the form

$$P_+ = \frac{m_1}{k_1} + \frac{m_2}{k_2} - \frac{\alpha B(\omega)}{r}, \tag{8.3}$$

$$P_- = m_1 k_1 + m_2 k_2, \tag{8.4}$$

$$K = -t(P_+ + P_-)/2 - \sum_{a=1}^{2}\frac{x_a m_a}{k_a} - $$
$$- \frac{\alpha}{r}\left[\left(\frac{x_1 k_2}{k_1} + \frac{x_2 k_1}{k_2}\right)f + \frac{1}{2}\left(\frac{k_1}{k_2} - \frac{k_2}{k_1}\right)\left(\frac{x_1 k_2}{k_1} - \frac{x_2 k_1}{k_2}\right)f'\right], \tag{8.5}$$

where

$$B(\omega) = 2\left(-\omega f + (\omega^2 - 1)f'\right), \tag{8.6}$$

permits one to reduce the solutions of Euler-Lagrange equations to quadrature [ 26]. But solutions of Euler-Lagrange equations exist only in the region $\mathcal{Q} \subset T\mathcal{M} \approx \mathbb{R}^4$ which is defined by the inequalities (3.5):

$$\mathsf{h}_f = \frac{m_1 m_2}{k_1^3 k_2^3} - \alpha\frac{(m_2 \kappa_2 + m_1 \kappa_1)A(\omega)}{r k_1^3 k_2^3} > 0, \tag{8.7}$$

where

$$A(\omega) = -f + \omega f' + (\omega^2 - 1)f''. \tag{8.8}$$

The investigation of two-particle models with the time-asymmetric field-like interactions (see [ 26, 25]) shows that for some values of the parameters the system reaches the boundary of the Lagrangian region $\partial\mathcal{Q} = \{(x_a, x_b, v_a, v_b) \in \mathbb{R}^4 | \mathsf{h}_\ell = 0; \mathsf{h}_\ell^{-1} = 0\}$. An exception is the repulsion case ($\alpha > 0$) if the total mass of the system $M > m_1 + m_2 = m$, where $m_1, m_2$ are particle rest masses. Then the system does not reach the singular points and the world lines are smooth timelike curves in $\mathbb{M}_2$ [ 46, 45]. The Hamiltonian description allows one to prolong the evolution of the system beyond the critical points for other values of the parameters and, as a result, to obtain continuous world lines in the following way [ 25].

The Legendre transformation $\mathscr{L}$ associated with the Lagrangian (8.1) with $f(\omega) = \omega^\ell$; $\ell = 0, 1, 2, ...$ has the form:

$$p_a = \frac{\partial L}{\partial v_a} = \frac{m_a}{k_a} + \frac{\alpha}{2r}\left(1 + \ell + (1-\ell)\frac{k_a^2}{k_a^2}\right)\omega^{\ell-1}. \tag{8.9}$$

Here $a = 1, 2$, $\bar{a} = 3 - a$.

In the scalar ($\ell = 0$) and vector ($\ell = 1$) cases it is possible to solve equations (8.9) with respect to velocities and obtain from the expressions for conserved quantities (3.3) the generators of the Lie algebra of the Poincaré group $\mathcal{P}(1,1)$ in the explicit form [ 59, 25]. Separation of the external and internal motions is carried out by the choice:

$$P_+ = p_1 + p_2 \ , \quad Q = K/P_+ \ ; \quad \{Q, P_+\} = 1 \tag{8.10}$$

as new external canonical variables. As internal variables we choose

$$\xi = \frac{m_2 p_1 - m_1 p_2}{P_+}, \quad q = r\frac{P_+}{m}; \quad \{q, \xi\} = 1, \tag{8.11}$$

where $m = m_1 + m_2$. Then the Hamiltonian equations of motion become

$$\dot{Q} = 1/2 - \frac{M^2}{2P_+^2}, \quad \dot{P}_+ = 0, \tag{8.12}$$

$$\dot{q} = \frac{1}{2P_+}\frac{\partial M^2}{\partial \xi}, \quad \dot{\xi} = -\frac{1}{2P_+}\frac{\partial M^2}{\partial q}. \tag{8.13}$$

Solving equations (8.9) with respect to velocities and substituting the solutions into the expression for the Hessian we obtain from (8.7) inequalities which define the image $\mathcal{LQ}$ of the Lagrangian region $\mathcal{Q}$ under the Legendre transformation (8.9). The external canonical variable $P_+$ is an integral of motion. The Hessian does not depend on the external variable $Q$. Thus, all the singularities of the Hessian are expressed in terms of inner variables and we can transform (8.7) into an inequality which defines the region $\widetilde{\mathcal{LQ}}$ in the inner phase space: $\widetilde{\mathcal{LQ}} \subset \mathbb{R}^2$. In the scalar case, if $\alpha > 0$, the region $\widetilde{\mathcal{LQ}}$ of the phase plane corresponds to the region $q > 0$ restricted by the curves $y_1$, $y_2$ (see figure 7) which are defined by the equations:

$$y_1 : -m_1\xi + m_1 m_2 + m_2\alpha/q = 0 \ ,$$
$$\tag{8.14}$$
$$y_2 : m_2\xi + m_1 m_2 + m_1\alpha/q = 0 \ .$$

If $\alpha < 0$, then $\widetilde{\mathcal{LQ}}$ lies between the curves $y_1$, $y_2$ to the right of their intersection point.

In the vector case, if $\alpha < 0$, the region $\widetilde{\mathcal{LQ}}$ corresponds to the region bounded by the curves $\tilde{y}_1$, $\tilde{y}_2$, $q = 0$ (see figure 8) which are defined by the equations:

$$\tilde{y}_1 : m_1 + \xi - \alpha/q = 0 \ ,$$
$$\tag{8.15}$$
$$\tilde{y}_2 : m_2 - \xi - \alpha/q = 0 \ .$$

If $\alpha > 0$, then the indicated region lies between the curves $\tilde{y}_1$, $\tilde{y}_2$ to the right of their intersection point. The intersection points of the phase trajectories and the curves $y_1, y_2$ ($\tilde{y}_1, \tilde{y}_2$) correspond to the case when one of the particles reaches the speed of light: $\kappa_1 = 0$ or $\kappa_2 = 0$.

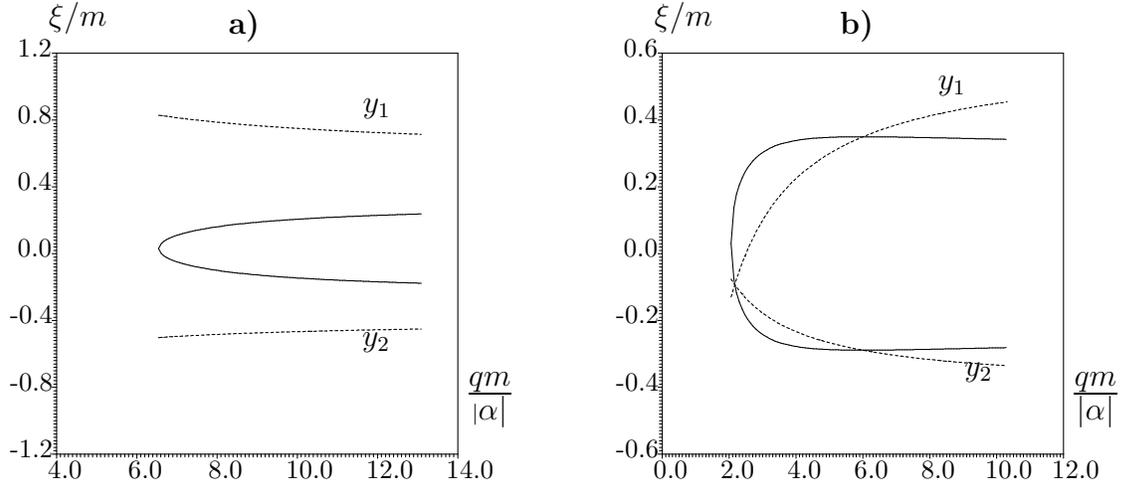

**Figure 7.** Scalar interaction. Phase trajectories (continuous curves): $(m_2-m_1)/m = 0.2$; $M/m = 1.2$. **a):** $\alpha > 0$, **b):** $\alpha < 0$. Dashed curves $y_1, y_2$ correspond to the singularity of the Hessian.

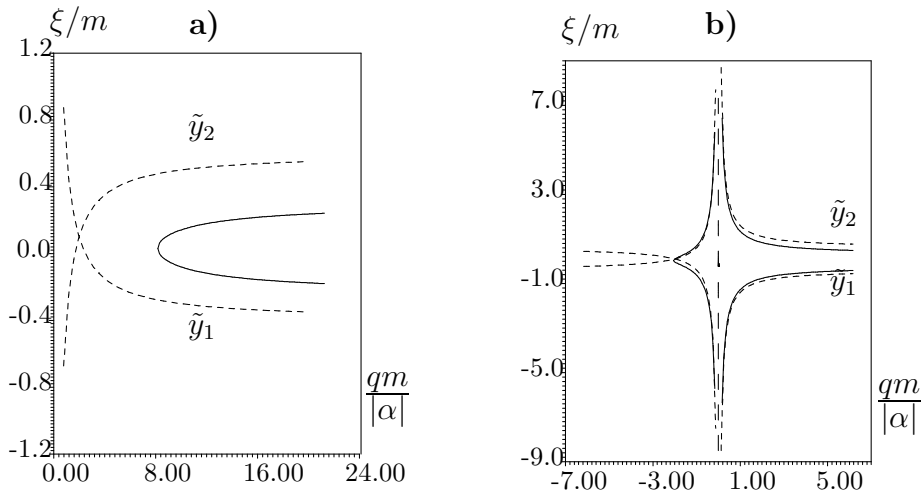

**Figure 8.** Vector interaction. Phase trajectories (continuous curves): $(m_2-m_1)/m = 0.2$; $M/m = 1.2$. **a):** $\alpha > 0$, **b):** $\alpha < 0$. Dashed curves $y_1, y_2$ correspond to the singularity of the Hessian.

To construct smooth world lines in $\mathbb{M}_2$ it is necessary to consider the inner motion in more detail. It is determined by the mass-shell equation

$$(\xi - \xi_M)^2 = \frac{(\nu^2 - 1)m^2 m_1^2 m_2^2 q^2 - 2\alpha M^2 m_1 m_2 m \nu^\ell q + (-1)^{\ell+1} M^4 \alpha^2}{M^4 q^2}, \tag{8.16}$$

where

$$\xi_M = \frac{(M^2 - m^2)(m_2 - m_2)}{2M^2}, \quad \nu = \frac{M^2 - m_1^2 - m_2^2}{2m_1 m_2} . \tag{8.17}$$

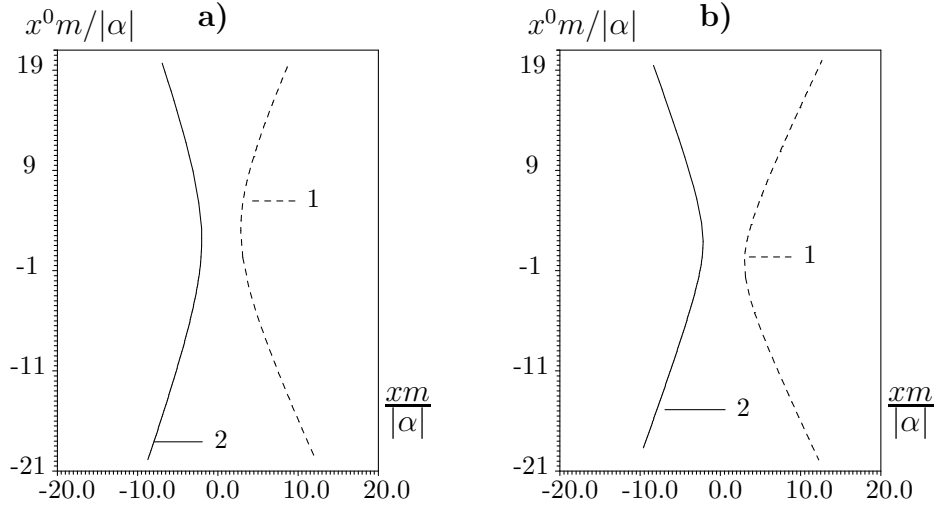

**Figure 9.** World lines in $\mathbb{M}_2$ for an unbounded motion: $(m_2 - m_1)/m = 0.2$, $M/m = 1.2$, $\alpha < 0$. **a)**: scalar interaction (Stephas case [ 46]); **b)**: vector interaction (Rudd and Hill case [ 45]).

We assume that equation (8.16) is true in the whole phase plane $\mathbb{R}^2$. The motion is possible in the region where

$$\mathsf{D}_\ell = (\nu^2 - 1)m^2 m_1^2 m_2^2 q^2 - 2\alpha M^2 m_1 m_2 m \nu^\ell q + (-1)^{\ell+1} M^4 \alpha^2 \tag{8.18}$$

is non-negative. Then, we see that for a bounded motion $q$ belongs to the interval $[q_1, q_2]$, where $q_1$, $q_2$ are real solutions of the quadratic equation $\mathsf{D}_\ell = 0$:

$$q_1 = \frac{2\alpha M^2 (-1)^{\ell+1}}{(M^2 - (m_1 - m_2)^2)m}, \quad q_2 = \frac{2\alpha M^2}{(M^2 - m^2)m}. \tag{8.19}$$

In such a manner we get the phase trajectories which lead to smooth world lines in $\mathbb{M}_2$ for all the values of the total mass of the system $M > 0$ and signs of the coupling constant $\alpha$ [ 25]. Using phase trajectory equation (8.16) and solving equations (8.12), (8.13) we obtain a parametric equation for world lines in $\mathbb{M}_2$:

$$x_1^0(q) = t(q) - x_1(q) , \quad x_2^0(q) = t(q) - x_2(q) ; \tag{8.20}$$

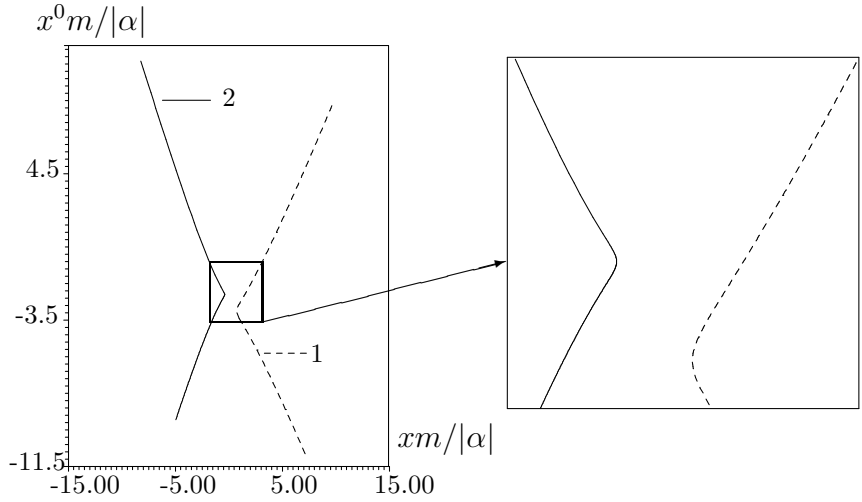

Figure 10. Scalar interaction. World lines in $\mathbb{M}_2$ for an unbounded motion: $(m_2 - m_1)/m = 0.2$, $M/m = 1.2$, $\alpha < 0$.

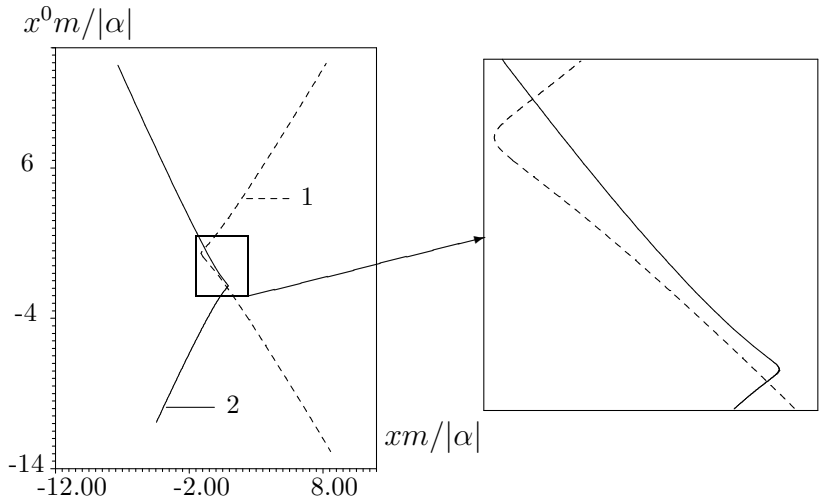

Figure 11. Vector interaction. World lines in $\mathbb{M}_2$ for an unbounded motion: $(m_2 - m_1)/m = 0.2$, $M/m = 1.2$, $\alpha < 0$.

$$x_1(q) = K/P_+ + \left(m_2 - \xi(M^2, q)\right) q/P_+ \ ,$$

$$x_2(q) = K/P_+ - \left(m_1 + \xi(M^2, q)\right) q/P_+ \ . \tag{8.21}$$

Figures 7,8 show examples of the phase trajectories for the scalar (figure 7) and vector (figure 8) interactions. Figures 9-11 show the corresponding smooth world lines.

Unlike the scalar interaction, there exist particle collisions in the vector case. At the collision points ($q = 0$) the particles mutually change their positions (figure 8, b) and the phase trajectories break up. The motion along smooth world lines corresponds to the jumps along the momentum axis $-\infty \to \infty$ ($\infty \to -\infty$).

# 9. Quantum models in $\mathbb{M}_2$

In this section we consider a number of exactly solvable quantum-mechanical models which follow from certain quantization procedures applied to the corresponding classical counterparts. We construct a quantum description for the investigated above classical time-asymmetric scalar and vector models, as well as for the classical models for which the Lagrangian description is not known.

## 9.1. Vector and scalar interactions

The classical two-particle system with time-asymmetric scalar and vector interactions can be quantized in a purely algebraic way [ 59] regarding the Lie algebra $\mathfrak{so}(2,1)$ as the basic algebraic structure. Let us introduce the following functions of canonical variables:

$$\begin{aligned}
J_0 &= \frac{1}{2}\left(\Delta q \xi^2 + \frac{q}{\Delta} + \frac{\alpha^2 \Delta(\alpha_0^2 - \alpha_1^2)}{q}\right), \\
J_1 &= \frac{1}{2}\left(\Delta q \xi^2 - \frac{q}{\Delta} + \frac{\alpha^2 \Delta(\alpha_0^2 - \alpha_1^2)}{q}\right), \\
J_2 &= q\xi \ ,
\end{aligned} \tag{9.1}$$

where $\Delta$ is an arbitrary constant. They span, under the Poisson bracketing, the Lie algebra $\mathfrak{so}(2,1)$

$$\{J_0, J_1\} = J_2, \qquad \{J_1, J_2\} = -J_0, \qquad \{J_2, J_0\} = J_1. \tag{9.2}$$

Then, the mass-shell equation (8.16) takes the form:

$$J + C_\ell = 0. \tag{9.3}$$

The quantity

$$J = aJ_0 + bJ_1 + dJ_2 \tag{9.4}$$

is an element of the Lie algebra of group $\mathcal{SO}(2,1)$ and we use the following notation:

$$a = \frac{M^2}{\Delta} + \Delta m_1 m_2 (m^2 - M^2), \quad b = \frac{M^2}{\Delta} - \Delta m_1 m_2 (m^2 - M^2),$$

$$d = (m_2 - m_1)(m^2 - M^2), \quad C_\ell = 2\alpha m m_1 m_2 \nu^\ell. \tag{9.5}$$

It would appear natural that the structure of the linear relation on the Lie algebra $\mathfrak{so}(2,1)$ must be preserved after quantization. Then, replacing functions (9.1) with the Hermitian operators obeying the commutation relations of the $\mathfrak{so}(2,1)$ Lie algebra

$$[\hat{J}_0, \hat{J}_1] = i\hat{J}_2, \qquad [\hat{J}_1, \hat{J}_2] = -i\hat{J}_0, \qquad [\hat{J}_2, \hat{J}_0] = i\hat{J}_1, \tag{9.6}$$

we obtain the quantum-mechanical equation:

$$(\hat{J} + C_\ell)|\psi\rangle = 0. \tag{9.7}$$

This equation was considered in [59] as the basic one for the quantum–mechanical problem. One can obtain in a purely algebraic way on the basis of equation (9.7) the mass spectrum

$$(M_n^\pm)_\ell^2 = m_1^2 + m_2^2 \pm 2m_1 m_2 (1 - (-1)^\ell \alpha^2 / n^2)^{(-1)^\ell/2}, \tag{9.8}$$

where

$$n = (-1 + \sqrt{1 + 4(-1)^\ell \alpha^2})/2 + s, \; s = 1, 2, \ldots \tag{9.9}$$

The branch $(M_n^+)_k^2$ has a correct non-relativistic limit. Expansion to the order $1/c^2$ gives the following correction to the energy spectrum:

$$E \approx -\frac{m_1 m_2 \alpha^2}{2m s^2 \hbar^2} - \frac{\alpha^4 m_1 m_2}{4m \hbar^4 s^4 c^2} \left[ \left( 1 - 4\ell + \frac{m_1 m_2}{m^2} \right) \frac{1}{2} - 4s(-1)^\ell \right],$$

$$s = 1, 2, \ldots . \tag{9.10}$$

In the single-particle limit $(m_1/m_2 \to 0)$ we obtain

$$E = m_1 \left( 1 - (-1)^\ell \alpha^2 / n^2 \right)^{(-1)^\ell/2} - m_1, \tag{9.11}$$

which is in agreement with a one-particle problem in the external scalar or vector field in the case of states with the zero value of the quantum orbital number. The mass spectrum of a vector type agrees with the result obtained by Barut on the basis of the infinite component wave equation [60].

The existence of an additional algebraic structure of the mass-shell equation permits one to quantize the classical problem without ambiguities typical of relativistic mechanics [61]. Furthermore, such a quantization method allows one to avoid difficulties connected with the choice of certain representation (coordinate, momentum, etc.) which is very important for the field-type interactions because of the difficulties of the global structure of the Hamiltonian description (see above).

## 9.2. Relativistic Hamiltonian models in $\mathbb{M}_2$

Considering the field-type models we started from the Lagrangian description. But it is also possible to construct a number of exactly solvable models immediately within the framework of the Hamiltonian description [ 62, 63, 64]. Contrary to the models based on the Fokker-type action integral, relativistic Hamiltonian models are not connected with the field theory. Nevertheless, they are also of interest for a variety of reasons. They can describe phenomenological aspects of the inner structure of mesons and baryons [ 65, 66]. Besides, these models can be useful for the verification of different approximation methods, and may be considered as an approximation of more realistic models. It appears to be significant for the explanation of relativistic effects in the well-established non-relativistic oscillator-like quark models of hadrons.

The standard quantization procedure consists in the transition from a set of canonical generators to a set of Hermitian operators which determine the unitary representation of the Poincaré group. So, in the case of two-dimensional space-time we must put in correspondence with the canonical generators of $\mathcal{P}(1,1)$ the Hermitian operators $\hat{K}, \hat{P}_+, \hat{P}_-$ in some Hilbert space which satisfy the following bracket relations:

$$[\hat{P}_+, \hat{P}_-] = 0, \qquad [\hat{K}, \hat{P}_\pm] = \pm i\hat{P}_\pm. \tag{9.12}$$

This determines the squared total mass operator $\hat{M}^2 = \hat{P}_+\hat{P}_-$ and the quantum problem is reduced to the eigenvalue problem [ 62, 63, 64]:

$$\hat{M}^2\psi = M_{n,\lambda}^2\psi. \tag{9.13}$$

From a variety of the known paths for such a transition we choose the Weyl quantization rule [ 67]. It is necessary that typical of the front form inequalities

$$p_a > 0 \tag{9.14}$$

be satisfied for this quantization method. It will be noted that these conditions are destroyed by field-like interactions. The wave functions $\psi(p) = \langle p|\psi\rangle$ describing the physical (normalized) states in the front form of dynamics constitute the Hilbert space $\mathcal{H}_N^F = \mathcal{L}^2(\mathbb{R}_+^N, d\mu_N^F)$ with the inner product [ 62, 63, 64]:

$$(\psi_1, \psi) = \int d\mu_N^F(p)\psi_1^*(p)\psi(p), \tag{9.15}$$

where

$$d\mu_N^F(p) = \prod_{a=1}^N \frac{dp_a}{2p_a}\Theta(p_a) \tag{9.16}$$

is a Poincaré–invariant measure and $\Theta(p_a)$ is the Heaviside function. According to the Weyl rule we get the following operators [ 62, 63, 64]:

$$\hat{P}_+ = \sum_{a=1}^N p_a, \quad \hat{K} = i\sum_{a=1}^N p_a\partial/\partial p_a, \quad \hat{P}_- = \hat{M}^2/\hat{P}_+ \quad, \tag{9.17}$$

which are Hermitian with respect to the inner product (9.15). They determine the unitary realization of group $\mathcal{P}(1.1)$ on the Hilbert space $\mathcal{H}_N^F$. Here $\hat{M}$ is determined by

$$\hat{M}^2 = \hat{M}_f^2 + \hat{V}, \tag{9.18}$$

where $\hat{M}_f^2$ is a free-particle part of the square mass operator:

$$\hat{M}_f^2 = \hat{P}_+ \sum_{a=1}^N \frac{m_a^2}{p_a}. \tag{9.19}$$

Operator $\hat{V}$ is an integral operator

$$(\hat{V}\psi)(p) = \int \mathrm{d}\mu_N^F(p) V(p,p') \psi(p') \tag{9.20}$$

with the kernel

$$V(p,p') = \left[ \prod_{d=1}^N \sqrt{4p_d p_d'} \right] \delta(P_+ - P_+') \int_{-\infty}^{\infty} V\left( r\frac{p_b + p_b'}{2}; \frac{r_{1c}}{r} \right) \times$$

$$\times \exp\left[ i \sum_{a=2}^N r_{1a}(p_a - p_a') \right] \prod_{a=2}^N \frac{\mathrm{d}r_{1a}}{2\pi}. \tag{9.21}$$

The general properties of the Weyl transformation [ 67 ] ensure that in the classical limit these operators correspond to the functions (3.7), (3.8).

The evolution of the quantum system is described in the front form of dynamics by the Schrödinger-type equation

$$\mathrm{i}\frac{\partial \Psi}{\partial t} = \hat{H}\Psi, \tag{9.22}$$

where $\Psi \in \mathcal{H}_N^F$ and

$$\hat{H} = \frac{1}{2}(\hat{P}_+ + \hat{P}_-) = \frac{1}{2}(\hat{P}_+ + \hat{M}^2/\hat{P}_+). \tag{9.23}$$

Putting $\Psi = \chi(t, P_+)\psi$, where $\psi$ is a function of some Poincaré-invariant inner variables, we obtain a stationary eigenvalue problem for the operator $\hat{M}^2$. In such a way a number of exactly solvable two-particle systems were considered in Refs. [ 62, 63 ]. It is convenient to introduce for a two-particle system the following Poincaré-invariant inner momentum variable [ 65 ]

$$\eta = (p_1 - p_2)/2P_+ , \tag{9.24}$$

which is linearly related to the variable $\xi = (m_2 - m_1)/2 + m\eta$. Then the interaction part of the squared total mass of the system $V$ takes the form:

$$V(rp_1, rp_2) = F(\rho, \eta) , \qquad \rho = rP_+ . \tag{9.25}$$

The conditions (9.14) lead to inequality $|\eta| < 1/2$ . The Hilbert space $\mathcal{H}_2^F$ decomposes into the tensor product $\mathcal{H}_2^F = h_{int} \otimes \mathcal{H}_{ext}^F$ , where "inner" and "external" spaces are realized, correspondingly, by functions $\psi(\eta)$ and $\chi(P_+)$ with the inner products

$$(\psi_1, \psi) = \frac{1}{2} \int\limits_{-1/2}^{1/2} \frac{\mathrm{d}\eta}{1/4 - \eta^2} \psi_1^*(\eta)\psi(\eta) \ , \qquad (9.26)$$

$$(\chi_1, \chi) = \int\limits_0^\infty \frac{\mathrm{d}P_+}{2P_+} \chi_1^*(P_+)\chi(P_+) \ . \qquad (9.27)$$

Operator $\hat{M}^2$ acts nontrivially only on $h_{int}$. It is an integral operator which is determined by the rule:

$$(\hat{M}^2\psi)(\eta) = \left( \frac{2m_1^2}{1+2\eta} + \frac{2m_2^2}{1-2\eta} \right) \psi(\eta) +$$

$$+ \int\limits_{-1/2}^{1/2} \mathrm{d}\eta' \sqrt{\frac{1-4\eta^2}{1-4\eta'^2}} W(\eta, \eta')\psi(\eta') \ , \qquad (9.28)$$

where kernel $W(\eta, \eta')$ has the form:

$$W(\eta, \eta') = \frac{1}{2\pi} \int\limits_{-\infty}^\infty \mathrm{d}\rho F\left( \rho, \frac{\eta + \eta'}{2} \right) \mathrm{e}^{-i\rho(\eta-\eta')} \ . \qquad (9.29)$$

The structure of operator $\hat{M}^2$ coincides with the one-dimensional variant of the corresponding expression in Ref. [ 65], but in the present treatment kernel $W(\eta, \eta')$ is directly related to the classical interaction potential $V$.

It is convenient to pass from the functions $\psi(\eta)$ with the inner product (9.26) to the functions

$$\varphi(\eta) = \frac{\psi(\eta)}{\sqrt{1/2 - 2\eta^2}} \qquad (9.30)$$

with the inner product

$$(\varphi_1, \varphi) = \int\limits_{-1/2}^{1/2} \mathrm{d}\eta \varphi_1^*(\eta)\varphi(\eta) \ . \qquad (9.31)$$

The latter differs from the non-relativistic product only by limits of integration. The action of $\hat{M}^2$ on the function $\varphi(\eta)$ is defined by the equation

$$(\hat{M}^2\varphi)(\eta) = \left( \frac{2m_1^2}{1+2\eta} + \frac{2m_2^2}{1-2\eta} \right) \varphi(\eta) + \int\limits_{-1/2}^{1/2} \mathrm{d}\eta' W(\eta, \eta')\varphi(\eta') \ . \qquad (9.32)$$

Let us consider two simple examples.

1. *δ–like potential* . Let us put $F(\rho,\eta) = \alpha\delta(\rho), \ \alpha = const$. Then the equation for $\varphi(\eta)$ has the form [ 62]:

$$\left(M^2 - \frac{m_1^2}{1/2+\eta} - \frac{m_2^2}{1/2-\eta}\right)\varphi(\eta) = \frac{\alpha}{2\pi}\int\limits_{-1/2}^{1/2}\mathrm{d}\eta'\varphi(\eta') \ . \qquad (9.33)$$

Putting

$$\int\limits_{-1/2}^{1/2}\mathrm{d}\eta\,\varphi(\eta) = C \ (\neq 0) \qquad (9.34)$$

we get from (9.33)

$$\varphi(\eta) = \frac{\alpha C}{2\pi}\left(M^2 - \frac{m_1^2}{1/2+\eta} - \frac{m_2^2}{1/2-\eta}\right)^{-1} \ . \qquad (9.35)$$

Substituting (9.35) into (9.34), we obtain the equation

$$\frac{2\pi}{\alpha} = \int\limits_{-1/2}^{1/2}\mathrm{d}\eta\left(M^2 - \frac{m_1^2}{1/2+\eta} - \frac{m_2^2}{1/2-\eta}\right)^{-1} \ , \qquad (9.36)$$

which describes the eigenvalues of $M^2$ for bound states.

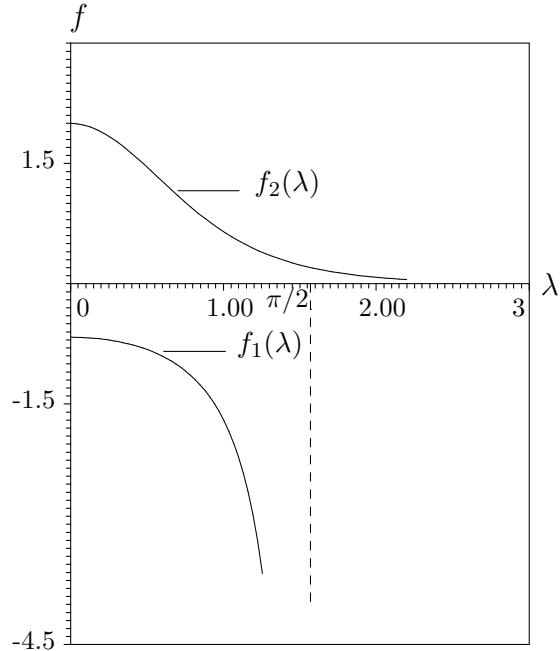

Figure 12. $\delta$-like potential. $f_1(\lambda)$ is a graph of the r.-h. side of equation (9.38), $f_2(\lambda)$ is a graph of the r.-h. side of equation (9.39)

Let us consider the case of equal particle masses ($m_1 = m_2 = m/2$). Then we get from (9.36)

$$\frac{2\pi M^2}{\alpha} = 1 - \frac{m^2}{2M} \int\limits_{-M}^{M} \frac{\mathrm{d}x}{x^2 + m^2 - M^2} \ . \tag{9.37}$$

If $M < m$, putting $M = m \sin\lambda, \ 0 \le \lambda \le \pi/2$, we come to the following transcendental equation for $\lambda$:

$$\frac{2\pi m^2}{\alpha} = \left(1 - \frac{2\lambda}{\sin 2\lambda}\right) \sin^{-2}\lambda \equiv f_1(\lambda) \ . \tag{9.38}$$

The graph of the right-hand side of this equation (figure 12) shows that there exists its only solution for $-3\pi m^2 < \alpha < 0$. This corresponds to attraction. The energy of a bound state has a proper non-relativistic limit.

It is interesting to point out that there also exists a bound state in the case of a strong repulsion. If $M > m$, one can put $M = m \operatorname{ch}\lambda, \ \lambda > 0$. Then, from (9.37) we obtain the following equation:

$$\frac{2\pi m^2}{\alpha} = \left(1 + \frac{2\lambda}{\operatorname{sh} 2\lambda}\right) \operatorname{ch}^{-2}\lambda \equiv f_2(\lambda) \ , \tag{9.39}$$

which has the only solution if $\alpha > \pi m^2$ (figure 12). This solution does not have a non-relativistic limit.

2. $\underline{Oscillator\ potential}$. Let us consider an interaction with a quadratical dependence on coordinates of the following type:

$$V = \omega_0^2 r^2 p_1 p_2 = \omega_0^2 (1/4 - \eta^2)\rho^2, \ \omega_0 \in \mathbb{R}. \tag{9.40}$$

Then, equation (9.13) transforms into an ordinary differential equation of the hypergeometric type [ 62, 63]:

$$\left(\frac{1}{4} - \eta^2\right)\varphi''(\eta) - 2\eta\varphi'(\eta) +$$
$$+ \left[-\frac{1}{2} + \frac{1}{\omega_0^2}\left(M_n^2 - \frac{m_1^2}{1/2 + \eta} - \frac{m_2^2}{1/2 - \eta}\right)\right]\varphi(\eta) = 0 \tag{9.41}$$

with the boundary conditions

$$\lim_{\eta \to \pm 1/2} u(\eta)\varphi(\eta) = 0, \ \lim_{\eta \to \pm 1/2} u(\eta)\varphi'(\eta) = 0 \ . \tag{9.42}$$

Equation (9.41) leads to the mass spectrum

$$M_n^2 = [m + \omega_0(n + 1/2)]^2 + \frac{\omega_0^2}{4}. \tag{9.43}$$

Its nontrivial solutions, which are bounded and square-integrable on the interval $(-1/2, \ 1/2)$ have the form:

$$\varphi_n(\eta) = C_n \left(\frac{1}{2} + \eta\right)^{m_1/\omega_0} \left(\frac{1}{2} - \eta\right)^{m_2/\omega_0} P_n^{(2m_2/\omega_0, 2m_1/\omega_0)}(2\eta) \ . \tag{9.44}$$

In equation (9.44) $P_n^{(2m_2/\omega_0, 2m_1/\omega_0)}(2\eta)$ are Jacobi polynomials [ 68] and $C_n$ are normalization constants.

In the non-relativistic limit $\hbar\omega_0/mc^2 \to 0$, $M_n \to m + E_n/c^2$ we obtain well–known wave functions in the momentum representation and a non-relativistic energy spectrum of the harmonic oscillator: $E = \hbar\omega_0(n + 1/2)$.

It is also possible to construct within the framework of the two-dimensional variant of the front form an exactly solvable quantum-mechanical $N$-particle model with the oscillator-like interaction

$$V = \omega_0^2 \sum_{a<b}\sum r_{ab}^2 p_a p_b. \qquad (9.45)$$

Function (9.45) gives an $N$-particle generalization of the two-particle interaction (9.40), as well as one of the possible relativistic generalizations of the N-particle oscillator potential. For this system by means of the Weyl quantization rule one can also reduce the eigenvalue problem (9.13) to a differential equation. The system with interaction (9.45) has $N-2$ additional integrals of motion which mutually commute and provide the exact integrability of the system in the classical case. They depend nontrivially on the products of coordinate and momenta variables [ 64]. Therefore, in general, the quantization procedure can destroy commutation relations between these quantities and, as a result, the integrability of the quantum problem. The Weyl quantization rule transforms classical additional integrals of motion into a set of quantum integrals of motion in involution. That permits one to solve exactly the eigenvalue problem and to obtain the eigenfunctions and eigenvalues of $\hat{M}^2$ (see [ 64]):

$$M_n^2 = \left[\sum_{a=1}^{N} m_a + \omega_0 \sum_{b=1}^{N-1}(n_b + 1/2)\right]^2 + \frac{N-1}{4}\omega_0^2. \qquad (9.46)$$

Interaction function (9.45) may be generalized by adding terms which are linear in the coordinates

$$V \to \tilde{V} = V + \alpha \sum_{a<b}\sum r_{ab}(p_a - p_b). \qquad (9.47)$$

Such a system also has additional integrals of motion and permits exact solutions in the quantum case [ 64].

Thus, the Weyl quantization rule preserves the commutation relation of Poincaré group $\mathcal{P}(1,1)$, as well as additional symmetries which are responsible for the integrability of this model [ 64]. As it was shown in [ 61] on the example of a two-particle oscillator-like model in the two-dimensional variant of the front form, the Weyl quantization is not the only quantization rule with this property. The application of different quantization rules preserving the commutation relation of $\mathcal{P}(1,1)$ may result in different observables as, for instance, a mass spectrum of the system [ 61].

## 10. Conclusion

We have considered the class of isotropic forms of dynamics which admit the construction of a variety of exactly solvable relativistic models of interacting particle systems. Most of the models originate from the Fokker-type actions with the time-asymmetric (retarded or advanced) Green function of the d'Alembert equation. These models reflect not only the relativistic kinematics but also certain field-theory aspects of the particle interaction. They demonstrate a complexity of the relativistic particle dynamics in comparison with its non-relativistic counterpart. The study of such a dynamics in detail is possible because of the fact that the considered forms of dynamics allow reformulation of the theory in terms of various formalisms and approaches, both three-dimensional and manifestly covariant four-dimensional.

The physical meaning of time-asymmetric interactions is not so clear. Nevertheless, the corresponding models may be regarded as the first step to some approximation scheme for finding solutions of more physically acceptable models, for example, the Wheeler-Feynman electrodynamics and the related theories. Particularly, in the linear approximation in the coupling constant the time-asymmetric, time-symmetric and purely retarded (field) approaches yield the same result. On the other hand, exact solutions of such models provide a better understanding of the special features of relativistic interactions and interrelations between various descriptions of relativistic interacting particles.

We wish to thank Yuriy Kluchkovsky, S. N. Sokolov, V. I. Lengel, and J. Llosa for the stimulating discussions. The ideas and influence of the late Professor Roman Gaida are evident throughout all the reported investigations.